\documentclass{emulateapj}

\newcommand{\etal}{et\,al.}

\newcommand{\gsim}{\raise0.3ex\hbox{$>$}\kern-0.75em{\lower0.65ex\hbox{$\sim$}}}

\newcommand{\lsim}{\raise0.3ex\hbox{$<$}\kern-0.75em{\lower0.65ex\hbox{$\sim$}}}

\newcommand{\msun}{M$_{\odot}$}
\newcommand{\mm}{$\mu$m}

\newcommand{\zsun}{Z$_{\odot}$}
\newcommand{\HI}{H\,{\sc i}}

\begin{document}     

\title{Dust and Atomic Gas in Dwarf Irregular Galaxies of the M\,81
  Group: The SINGS and THINGS view.}
\author{Fabian Walter\altaffilmark{1},
John M. Cannon\altaffilmark{1,2},
H{\'e}l{\`e}ne Roussel\altaffilmark{1},
George J. Bendo\altaffilmark{3},
Daniela Calzetti\altaffilmark{4},
Daniel A. Dale\altaffilmark{5},
Bruce T. Draine\altaffilmark{6},
George Helou\altaffilmark{7},
Robert C. Kennicutt, Jr.\altaffilmark{8,9},
John Moustakas\altaffilmark{9,10},
George H. Rieke\altaffilmark{9},
Lee Armus\altaffilmark{7},
Charles W. Engelbracht\altaffilmark{9},
Karl Gordon\altaffilmark{9},
David J. Hollenbach\altaffilmark{11},
Janice Lee\altaffilmark{9},
Aigen Li\altaffilmark{12},
Martin J. Meyer\altaffilmark{4},
Eric J. Murphy\altaffilmark{13},
Michael W. Regan\altaffilmark{4},
John-David T. Smith\altaffilmark{9},
Elias Brinks\altaffilmark{14},
W.~J.~G. de Blok\altaffilmark{15},
Frank Bigiel\altaffilmark{1},
Michele D. Thornley\altaffilmark{16}
}
\altaffiltext{1}{Max-Planck-Institut f{\"u}r Astronomie,
  K{\"o}nigstuhl 17, D-69117, Heidelberg, Germany; walter@mpia.de} 
\altaffiltext{2}{Astronomy Department, Wesleyan University,
  Middletown, CT 06459}
\altaffiltext{3}{Imperial College London, Prince Consort Road, London
  SW7 2AZ United Kingdom}
\altaffiltext{4}{Space Telescope Science Institute, 3700 San Martin
  Drive, Baltimore, MD 21218}
\altaffiltext{5}{Department of Physics and Astronomy, University of
  Wyoming, Laramie, WY 82071}
\altaffiltext{6}{Princeton University Observatory, Peyton Hall,
  Princeton, NJ 08544}
\altaffiltext{7}{California Institute of Technology, MC 314-6,
  Pasadena, CA 91101; gxh@ipac.caltech.edu}
\altaffiltext{8}{Institute of Astronomy, University of Cambridge,
  Madingley Road, Cambridge CB3 0HA, UK}
\altaffiltext{9}{Steward Observatory, University of Arizona, 933 North
  Cherry Avenue, Tucson, AZ 85721} 
\altaffiltext{10}{Department of Physics, New York University, 4 Washington Place, New York, NY 10003}
\altaffiltext{11}{NASA/Ames Research Center, MS 245-6, Moffett Field, CA,
94035}
\altaffiltext{12}{Department of Physics and Astronomy, University of
Missouri, Columbia, MO 65211}
\altaffiltext{13}{Department of Astronomy, Yale University, New Haven, CT 
  06520}
\altaffiltext{14}{Centre for Astrophysics Research, University of
  Hertfordshire, Hatfield AL10 9AB, U.K.}
\altaffiltext{15}{Research School of Astronomy \& Astrophysics, Mount
  Stromlo Observatory, Cotter Road, Weston ACT 2611, Australia}
\altaffiltext{16}{Department of Physics, Bucknell University, Lewisburg, PA
  17837}

\begin{abstract}
  We present observations of the dust and atomic gas phase in seven
  dwarf irregular galaxies of the M\,81 group.  The far--infrared data
  have been obtained as part of the `Spitzer Infrared Nearby Galaxies
  Survey {\it SINGS}'. Maps of the distribution of atomic hydrogen
  (HI) have been obtained through `The \HI\ Nearby Galaxy Survey {\it
  THINGS}'. The Spitzer observations provide a first glimpse of the
  nature of the non--atomic ISM in these metal--poor
  (Z$\sim$0.1\,Z$_\odot$), quiescent
  (SFR$\sim$0.001--0.1\,M$_\odot$\,yr$^{-1}$) dwarf galaxies.  Dust
  emission is detected in five out of the seven targets (the two
  systems with the lowest star formation rates are non--detections).
  Most detected dust emission is restricted to \HI\ column densities
  $>1\times10^{21}$\,cm$^{-2}$ and almost all regions of high \HI\
  column density ($>2.5\times10^{21}$\,cm$^{-2}$) have associated dust
  emission.  Spitzer spectroscopy of two regions in the brightest
  galaxies (IC~2574 and Holmberg~II) show distinctly different
  spectral shapes.  The spectrum of IC~2574 shows aromatic features
  that are less luminous (relative to the FIR luminosity) compared to
  an average SINGS spiral galaxy by a factor ot $\sim7$ .  The
  aromatic features in Holmberg~II (which has only a slightly lower
  gas--phase metallicity) are fainter than in IC~2574 by an order of
  magnitude.  This result emphazises that the strength of the aromatic
  features is not a simple linear function of metallicity.  Whereas
  the \HI\ masses are well-constrained, model dependencies make it
  difficult to measure the dust masses with a high degree of
  confidence. We estimate dust masses of $\sim$10$^4$--$10^{6}$ \msun
  for the M\,81 dwarf galaxies, resulting in an average dust--to--gas
  ratio (M$_{\rm dust}$/M$_{\rm \HI}$) of $\sim3\times10^{-4}$
  ($1.5\times 10^{-3}$ if only the \HI\ that is associated with dust
  emission is considered); this is an order of magnitude lower than
  the typical value derived for the SINGS spirals.  The dwarf galaxies
  are underluminous per unit star formation rate at 70\mm\ as compared
  to the more massive galaxies in {\em SINGS} by a factor of $\sim 2$.
  However, the average 70\,\mm/160\,\mm\ ratio in the sample dwarf
  galaxies is higher than what is found in the other galaxies of the
  SINGS sample.  This can be explained by a combination of a lower
  dust content in conjunction with a higher dust temperature in the
  dwarfs (likely due to the harder radiation fields in the low
  metallicity environments).
\end{abstract}                                          


\section{Introduction}

Nearby dwarf galaxies have proven to be ideal laboratories to
investigate how stars form out of gas and how, in turn, violent star
formation shapes the ambient interstellar medium (ISM). These systems
are highly susceptible to the formation of shells and holes in the
neutral gas phase; more energetic star formation can lead to the
formation of gaseous outflows from these systems. Previous studies
have shown that dwarf galaxies can be used as testbeds of `simple'
prescriptions for star formation (since these systems are typically in
solid--body rotation and are therefore less affected by shear in the
ISM) and for understanding the connection between mechanical energy
input into the ISM (`feedback') and future star formation.  Given
their low metallicities, they also provide a unique opportunity to
study the conditions of the ISM in environments that may resemble
those in the earliest starforming systems at high redshift.

\begin{deluxetable*}{lcccccccccc} 
  \tabletypesize{\scriptsize} \tablecaption{Properties of the sample
    dwarf galaxies in the M\,81 Group} \tablewidth{0pt} \tablehead{
    \colhead{Galaxy} &\colhead{D$^{\rm a}$} & \colhead{F$_{\rm HI}$}
    &\colhead{M$_{\rm HI}$}
    &\colhead{log(F(H$\alpha$))\tablenotemark{b}}
    &\colhead{12+log(O/H)\tablenotemark{c}} &\colhead{S(24\,\mm)\tablenotemark{d}}&\colhead{S(70\,\mm)\tablenotemark{d}}&\colhead{S(160\,\mm)\tablenotemark{d}}\\
    \colhead{} &\colhead{(Mpc)} &\colhead{(Jy\,km\,s$^{-1}$)}
    &\colhead{(10$^8$ M$_{\odot}$)}
    &\colhead{(erg\,s$^{-1}$\,cm$^{-2}$)} &\colhead{}
    &\colhead{(Jy)} & \colhead{(Jy)} & \colhead{(Jy)} } \startdata
  IC~2574     &4.02    & 386.7 &14.75  &  -11.27   & 7.94$\pm$0.06    & 0.28$\pm$0.013  & 5.55$\pm$0.42 & 11.75$\pm$1.50 \\
  Holmberg~II &3.39    & 219.3 &5.95   &  -11.30   & 7.68$\pm$0.03    & 0.20$\pm$0.008  & 3.67$\pm$0.26 & 4.46$\pm$0.58 \\
  Holmberg~I  &3.84    & 40.1  &1.40   &  -12.43   & 7.54$\pm$0.34    & 0.013$\pm$0.002 & 0.41$\pm$0.08 & 0.90$\pm$0.17  \\
  DDO\,165     &4.57    & 35.0  &1.72   &  -12.93   & 7.76$\pm$0.18$^d$& $<$0.014        & $<$0.15        & $<$0.33      \\
  DDO\,053     &3.56    & 20.0  &0.60   &  -12.24   & 7.77$\pm$0.1     & 0.029$\pm$0.001 & 0.40$\pm$0.03 & 0.50$\pm$0.11 \\
  M\,81\,DwB   &5.3     & 3.8   &0.25   &  -12.82   & 7.85$\pm$0.17    & 0.009$\pm$0.001 & 0.15$\pm$0.03 & 0.39$\pm$0.18 \\
  M\,81\,DwA   &3.55    & 4.1   &0.12   & --        & --               & $<$0.002        & $<$0.17       & $<$0.15
  \enddata 
\tablenotetext{a}{Distances are from Karachentsev et al.\ 2002, 2003} 
\tablenotetext{b}{H$\alpha$ fluxes are from Kennicutt et al.\ 2006, Lee 2006}
\tablenotetext{c}{Characteristic gas phase oxygen abundances from
Moustakas et al.\ 2007, based on the Pilyugin \& Thuan 2005
strong--line calibration of R$_{23}$=([OII]+[OIII])/H$\beta$; 12+log(O/H)$_\odot$=8.7 (Asplund et al.\ 2004).}
  \tablenotetext{d}{Flux densities adopted from Dale et al.\ 2006.}
\tablenotetext{d}{Based on the luminosity--metallicity relation, Moustakas et al.\ 2007}
\label{t1}
\end{deluxetable*}

Based on their ISM properties, dwarf galaxies can be roughly divided
into two subgroups: gas-rich dwarf irregulars (dIrrs) and gas-poor
dwarf spheroidals/ellipticals (dSph/dE).  The gas-rich dIrrs are
particularly interesting for studies of current star formation, since
they still contain the `fuel' for star formation. Though many studies
have shown that these objects are rich in atomic hydrogen (\HI; for a
review see {Skillman 1996}\nocite{skillman96conf}), little is known
about their molecular gas properties. Many searches for molecular gas
(through observations of the most abundant tracer molecule, CO) have
been performed but very few dwarf galaxies have been detected in
CO so far: no dwarf has been detected in CO at metallicities \lsim\ 
10\% \zsun\ \citep{taylor98,barone00,leroy05}.

Similarly, little is known about the dust properties in faint, low
metallicity dwarf galaxies. Some of the brightest dwarfs have been
detected with previous far-infrared (far-IR) observatories such as
{\it IRAS} and {\it ISO} (see, for example, {Hunter \etal\
  1989}\nocite{hunter89}, {Gallagher \etal\ 1991}\nocite{gallagher91},
{Melisse \& Israel 1994a,b}\nocite{melisse94a,melisse94b}, {Hunter
  \etal\ 2001}\nocite{hunter01}).  These early studies found that
dwarfs typically have higher dust temperatures than those derived for
more massive galaxies, with the peak of the infrared spectral energy
distribution (SED) shifted to shorter wavelengths.

The more typical `quiescent' dIrr galaxies (with
SFR$\leq$0.1\,M$_\odot$\,yr$^{-1}$) had remained undetected in the
far--IR prior to the advent of the {\it Spitzer Space Telescope}; its
dramatically increased sensitivity compared to previous observatories
has opened up the low-metallicity regime of the extragalactic ISM to
exploration in the far-IR.  Early {\it Spitzer} observations have
naturally concentrated on the brightest and most extremely metal-poor
dwarfs.  \citet{houck04} show that SBS\,0335$-$052, one of the most
metal-poor dwarf galaxies known, has an exceptional SED that is
shifted blueward with a peak at $\sim$ 28\,\mm\ (compared to
$\sim$80~\mm\ for more metal-rich starburst galaxies and $>$100~\mm\
for local star-forming spirals).
\citet{engelbracht04} presented observations of the dwarf galaxy
NGC\,55, and a larger sample of galaxies showed that there is a
metallicity threshold above which emission from aromatic features
appear \citep{engelbracht05}. Smith et al. (2006) used {\it Spitzer}
spectroscopy to study how the spectral signatures of aromatic features
change as a function of metallicity. Other recent {\it Spitzer}
studies of actively star forming dwarf galaxies include spectroscopic
observations by Wu et al.\ 2006 and O'Halloren et al.\ 2006 and
imaging studies of star--forming dwarfs and Local Group galaxies
(including dwarf galaxies) by Rosenberg et al.\ 2006 and Jackson et
al.\ 2006.  {Cannon \etal\ 2005,
2006a,b}\nocite{cannon05,cannon06a,cannon06b} have presented detailed
{\it Spitzer} studies of some of the most luminous and nearby dwarfs
(IC~2574, NGC\,1705, and NGC\,6822).  These case studies have shown
that localized star formation has dramatic effects on the multiphase
ISM, altering the relative strengths of nebular, far-IR dust, and
radio continuum emission.

Here we discuss the distribution of dust and atomic hydrogen in seven
dwarf irregular galaxies in the M\,81 group of galaxies.  Our sample
spans two orders of magnitude in \HI\ masses and a similar range in star
formation rates, from $<0.001$\,M$_\odot$\, yr$^{-1}$ (H$\alpha$
non--detections) to $\sim$0.1\,M$_\odot$\, yr$^{-1}$.  The general
properties of the sample dwarfs are summarized in Table~1.  All these
dwarf irregular galaxies are part of the ` Spitzer Infrared Nearby
Galaxies Survey' ({\it SINGS}, see {Kennicutt \etal\
2003}\nocite{kennicutt03}) and have also been included in `The \HI\
Nearby Galaxy Survey' ({\it THINGS}, {Walter \etal\
2005}\nocite{walter05conf}).

This paper is organized as follows: in \S~\ref{S2} we summarize the
{\it Spitzer} MIPS and the VLA \HI\ observations; in \S~\ref{S3}
we present our results, i.e. the distribution of the atomic gas and
the dust, IRS spectra of two galaxies, estimates of the dust masses
and a comparison to other galaxies in the {\it SINGS} sample.  In
\S~\ref{S4} we present a summary of our study.

\begin{figure*}
\epsscale{0.8}
\plotone{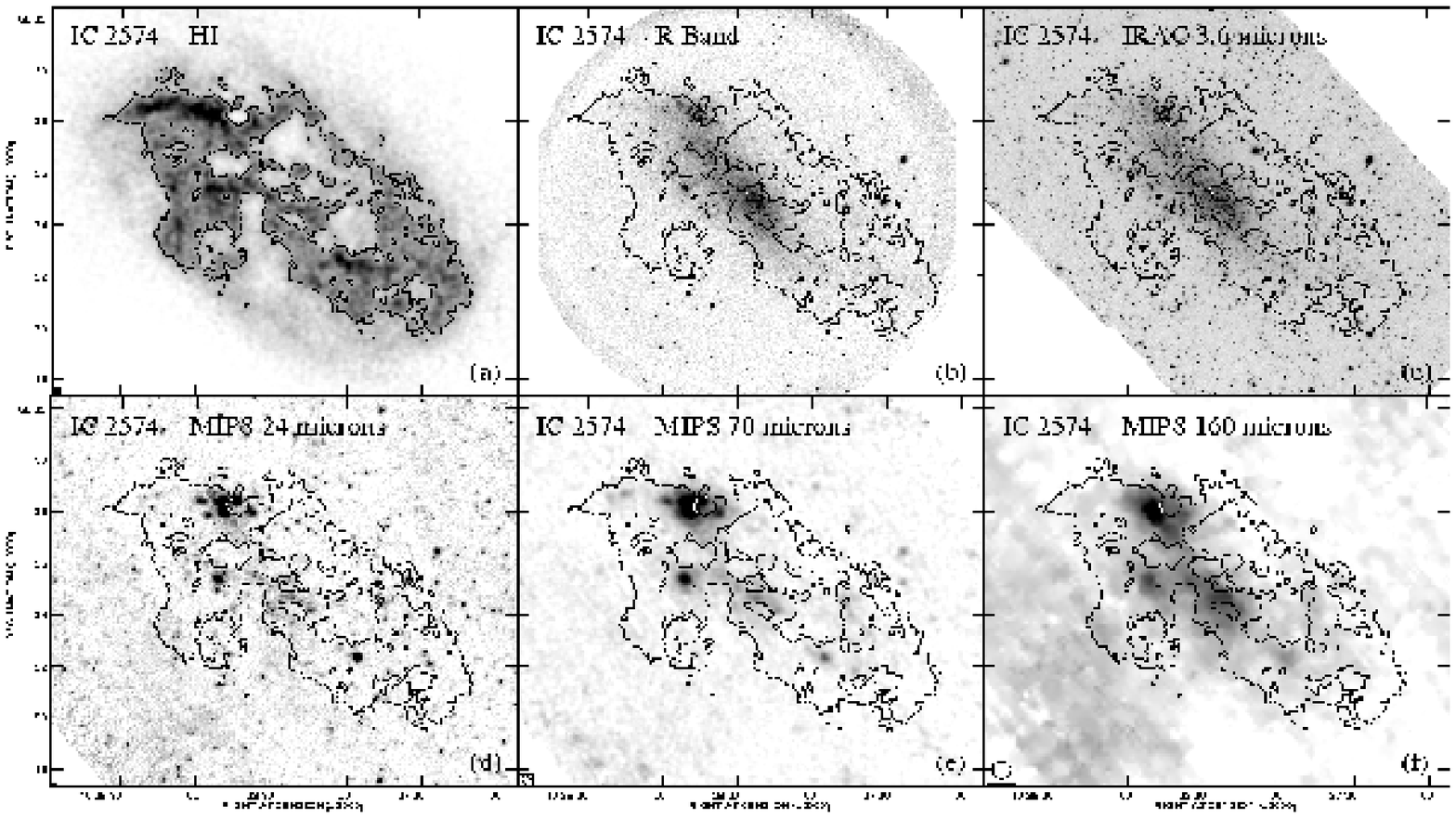}
\caption{IC~2574: (a) integrated {\em THINGS} \HI\ map (contour shown
  at N$_{\rm HI}$=$10^{21}\,$cm$^{-2}$ in all panels); (b) R--band
  image; (c) IRAC 3.6\mm\ image; (d), (e), and (f): 24\,\mm, 70\,\mm\
  and 160\,\mm\ images, respectively. The sizes of the \HI\ and MIPS
  beams are given in the lower left corners of their respective panels.}
\label{figcap1}
\end{figure*}

\begin{figure*}
\epsscale{0.8} 
\plotone{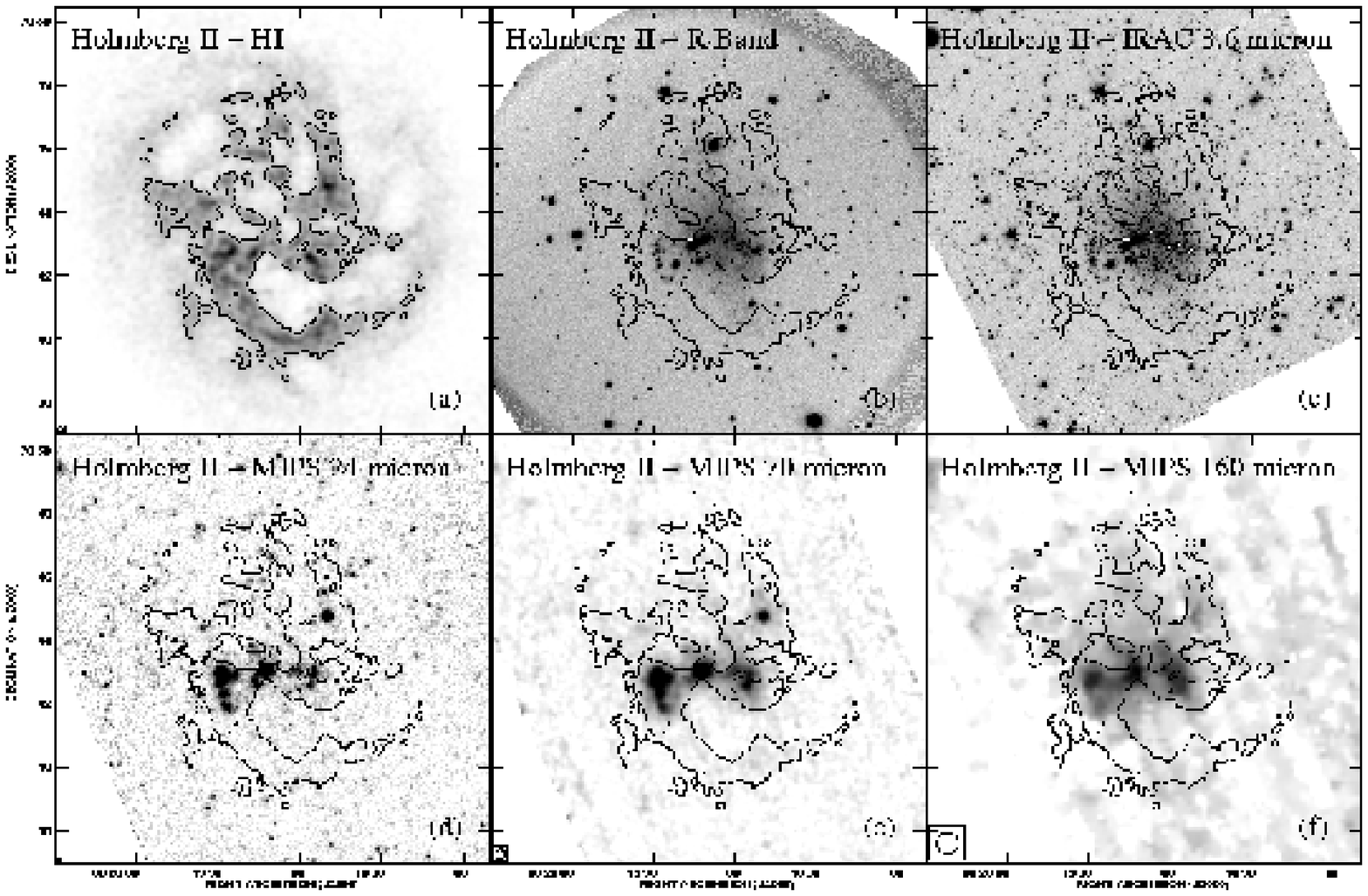}
\caption{Holmberg~II: (a) integrated {\em THINGS} \HI\ map (contour
  shown at N$_{\rm HI}$=$10^{21}\,$cm$^{-2}$ in all panels); (b)
  R--band image; (c) 3.6\mm\ image; (d), (e), and (f): 24\,\mm,
  70\,\mm\ and 160\,\mm\ images, respectively.  The sizes of the \HI\
  and MIPS beams are given in the lower left corners of their respective panels.}
\label{figcap2}
\end{figure*}

\section{Observations}
\label{S2}
\subsection{{\it Spitzer} MIPS Observations}
\label{S2.1}

The 24, 70, and 160\,\mm\ data were taken with MIPS on the Spitzer
Space Telescope as part of the {\it SINGS} survey (Kennicutt et al.\
2003).  The observations were obtained using the scan-mapping mode in
two separate visits to each galaxy (facilitating removal of asteroids
and detector artifacts). Each pixel in the map was observed 40, 20,
and 4 times at 24, 70, and 160\mm, respectively, resulting in
integration times per pixel of 160, 80, and 16~s, respectively.  All
MIPS data were processed using the MIPS Instrument Team Data Analysis
Tool \citep{gordon05}.  Systematic uncertainties (e.g., detector
nonlinearities, time-dependent responsivity variations, background
removal, etc.)  limit the absolute flux calibration to $\sim$4\%, 7\%
and 12\% in the MIPS 24\,\mm, 70\,\mm\ and 160\,\mm\ bands.  The FWHM
of the MIPS PSFs are 6\arcsec, 18\arcsec, and 40\arcsec\ at 24\mm,
70\mm\ and 160\,\mm, respectively.  For more details on the MIPS data
reduction see Bendo et al.\ (2006) -- for a general description of the
{\it SINGS} observing strategies see Kennicutt et al.\
(2003)\nocite{kennicutt03}.

The flux densities presented in this paper were derived for apertures
much larger than the MIPS PSFs, and therefore aperture corrections
have not been applied.  The M\,81 group is located in a direction
where the Galaxy is rich in infrared cirrus \citep{devries87}.
Inspection of individual images reveals that Galactic cirrus emission
is indeed present in the longer wavelength MIPS images. However, this
emission is distributed over much larger angular scales than the
sources of interest and can thus be easily separated from the galaxies
presented in this study.

The aperture for each individual galaxy has been chosen carefully to
encompass all the emission visible in all three MIPS bands; apertures
were compared with IRAC band 1 and \HI\ imaging to ensure that the total
galaxy extent (i.e., gas and stars) was measured. For each individual
galaxy, we used the same aperture to extract the flux densities from
the three MIPS bands. To account for the variations in the background
(both instrumental and due to Galactic Cirrus) we have defined
multiple background regions for each galaxy that contain the same area
as the target aperture. The (background subtracted) source flux
densities we derived using this technique are the same (within the
errors) as the values derived by Dale et al.\ 2005, 2006 in their
study of the entire SINGS sample. For consistency we therefore adopt
the values of Dale et al.\ 2006 for our study. The global flux
densities are summarized in Table~1 --- note that M\,81\,dwA and
DDO\,165 are MIPS non--detections. The reader is referred to Dale et
al.\ 2006 for the IRAC flux densities of the dwarf galaxies in our sample.

\begin{figure*}
\epsscale{0.8}
\plotone{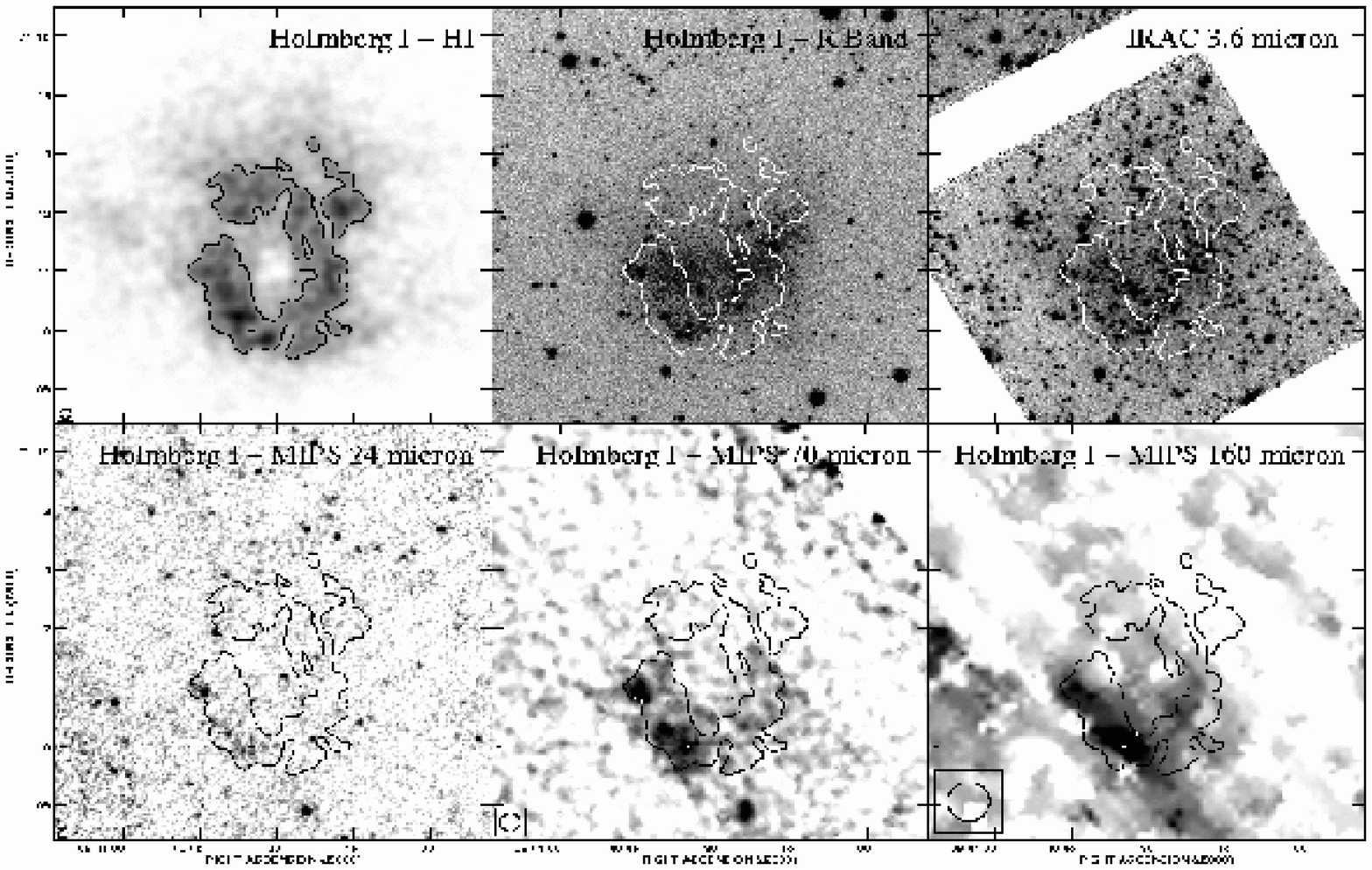}
\caption{Holmberg~I: (a) integrated {\em THINGS} \HI\ map (contour
  shown at N$_{\rm HI}$=$10^{21}\,$cm$^{-2}$ in all panels); (b)
  R--band image; (c) IRAC 3.6\mm\ image; (d), (e), and (f): 24\,\mm,
  70\,\mm\ and 160\,\mm\ images, respectively.  The sizes of the \HI\
  and MIPS beams are given in the lower left corners of their respective panels.}
\label{figcap3}
\end{figure*}

\begin{figure*}
\epsscale{0.8}
\plotone{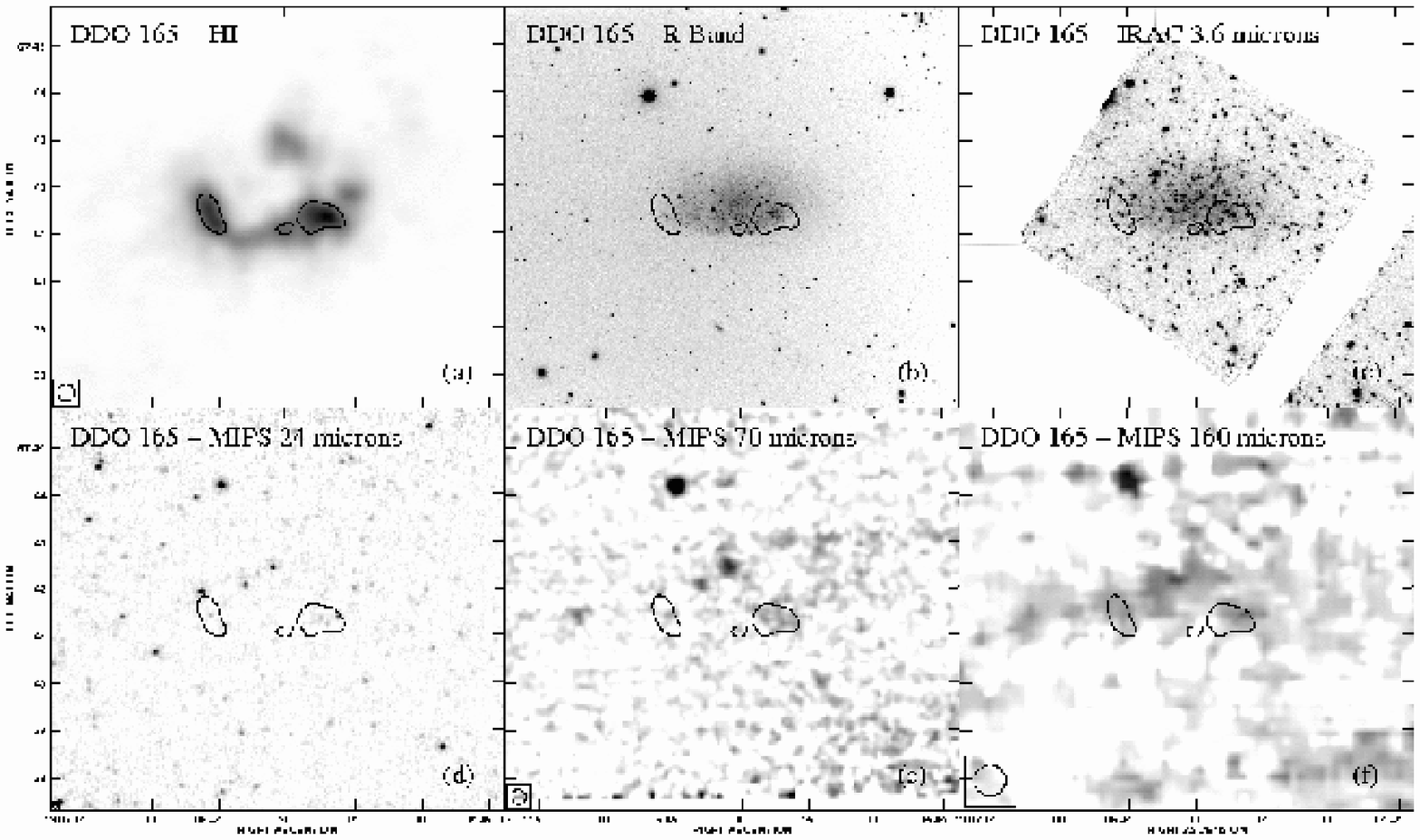}
\caption{DDO\,165: (a) integrated \HI\ map at 21$''$ resolution (data
  from Cannon et al.\ in prep.; contour shown at N$_{\rm
    HI}$=$10^{21}\,$cm$^{-2}$ in all panels); (b) R--band image; (c)
  IRAC 3.6\mm\ image; (d), (e), and (f): 24\,\mm, 70 \,\mm\ and
  160\,\mm\ images, respectively.  The sizes of the \HI\ and MIPS
  beams are given in the lower left corners of their respective
  panels.}
\label{figcap4}
\end{figure*}

\subsection{VLA Observations}
\label{S2.2}

\HI\ data for six of the seven M\,81 group dwarfs presented here were
obtained as part of `The \HI\ Nearby Galaxy Survey' ({\it THINGS}), a
survey to obtain high-resolution NRAO\footnote{The National Radio
  Astronomy Observatory is a facility of the National Science
  Foundation operated under cooperative agreement by Associated
  Universities, Inc.} VLA \HI\ imaging for 35 nearby galaxies
\citep{walter05conf}. \HI\ data for DDO\,165 were taken from Cannon et
al.\ (in prep.).

For {\it THINGS}, each galaxy was observed with the {\it VLA} in D, C
and B configurations with typical integration times of 1.5 hours, 2.5
hours and 7 hours, respectively.  The calibration and data reduction
were done using the {\sc AIPS} package\footnote{The Astronomical Image
  Processing System ({\sc AIPS}) has been developed by the NRAO.}.
The absolute flux scale for the data was determined by observing the
quasar 3C286 in all observing runs.  The time variable phase and
amplitude calibration were done using the nearby, secondary
calibrators 1313$+$549 and 1252$+$565 which are unresolved for the
arrays used.  The {\it uv}-data were inspected for each array and bad
data points due to either interference or cross-talk between antennae
were removed, after which the data were calibrated.  After final
editing, all data for each target were combined to form a single
dataset which was subsequently used to create maps of the brightness
distribution on the sky as a function of frequency/velocity (data
cubes).

In order to remove the continuum from the line data we first
determined the line-free channels in our observations and subtracted
the continuum emission in the {\it uv}-plane. After that, datacubes
(1024 $\times$ 1024 pixels $\times$ 80 channels each) were produced
using the task {\sc imagr} in {\sc AIPS}. To boost the angular
resolution while still maintaining a reasonable noise, we use a {\sc
robust} parameter of 0.5 for the final imaging. To ensure that we
reach identical beam sizes for all {\it THINGS} observations, the data
were subsequently convolved to a common resolution of 10$"$.  This
resulted in a typical rms noise per channel of 0.5 mJy\,beam$^{-1}$
for a 2.5\,km s$^{-1}$ channel (corresponding to
N$_{\rm HI}$=1.5$\times$10$^{19}$\,cm$^{-2}$).  To separate real emission from
noise in the final integrated \HI\ maps, we only consider regions
which show emission in consecutive channels above a set level
($\sim2\sigma$) in slightly convolved (20$''$) cubes. Note that the
data for DDO\,165 are at a resolution of 21\arcsec.

The fluxes in the integrated {\it THINGS} \HI\ map are corrected for
the fact that typically the residual flux of the source in cleaned
channel maps is overestimated (sometimes by a factor of a few) due to
the different beam sizes of the dirty and cleaned beams (for details
see, e.g., {J\"ors\"ater \& van Moorsel 1995}\nocite{jorsater95},
{Walter \& Brinks 1999}\nocite{walter99}). In the integrated \HI\ maps
this typically leads to flux corrections of order 25--40\%. To correct
for this, we have scaled the residual fluxes by the ratio of the dirty
and clean beam sizes and estimate that our column densities are
correct within 10\% (including the intrinsic uncertainties of the flux
calibration; the inferred \HI\ masses are given in Table~1).

\begin{figure*}
\epsscale{0.8}
\plotone{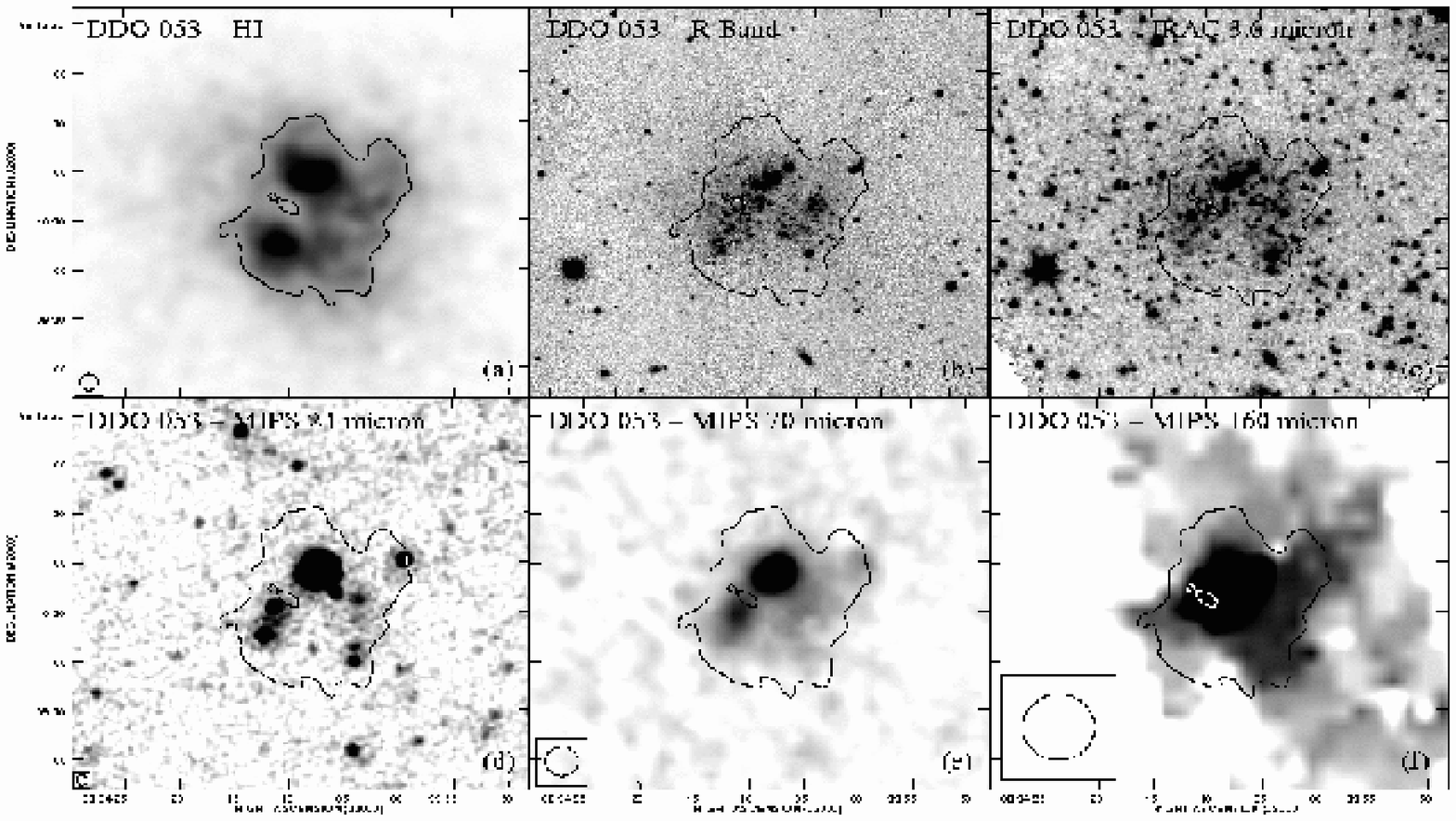}
\caption{DDO\,53: (a) integrated {\em THINGS} \HI\ map (contour shown
  at N$_{\rm HI}$=$10^{21}\,$cm$^{-2}$ in all panels); (b) R--band
  image; (c) IRAC 3.6\mm\ image; (d), (e), and (f): 24\,\mm, 70\,\mm\
  and 160\,\mm\ images, respectively. The sizes of the \HI\ and MIPS
  beams are given in the lower left corners of their respective panels.}
\label{figcap5}
\end{figure*}

\begin{figure*}
\epsscale{0.8}
\plotone{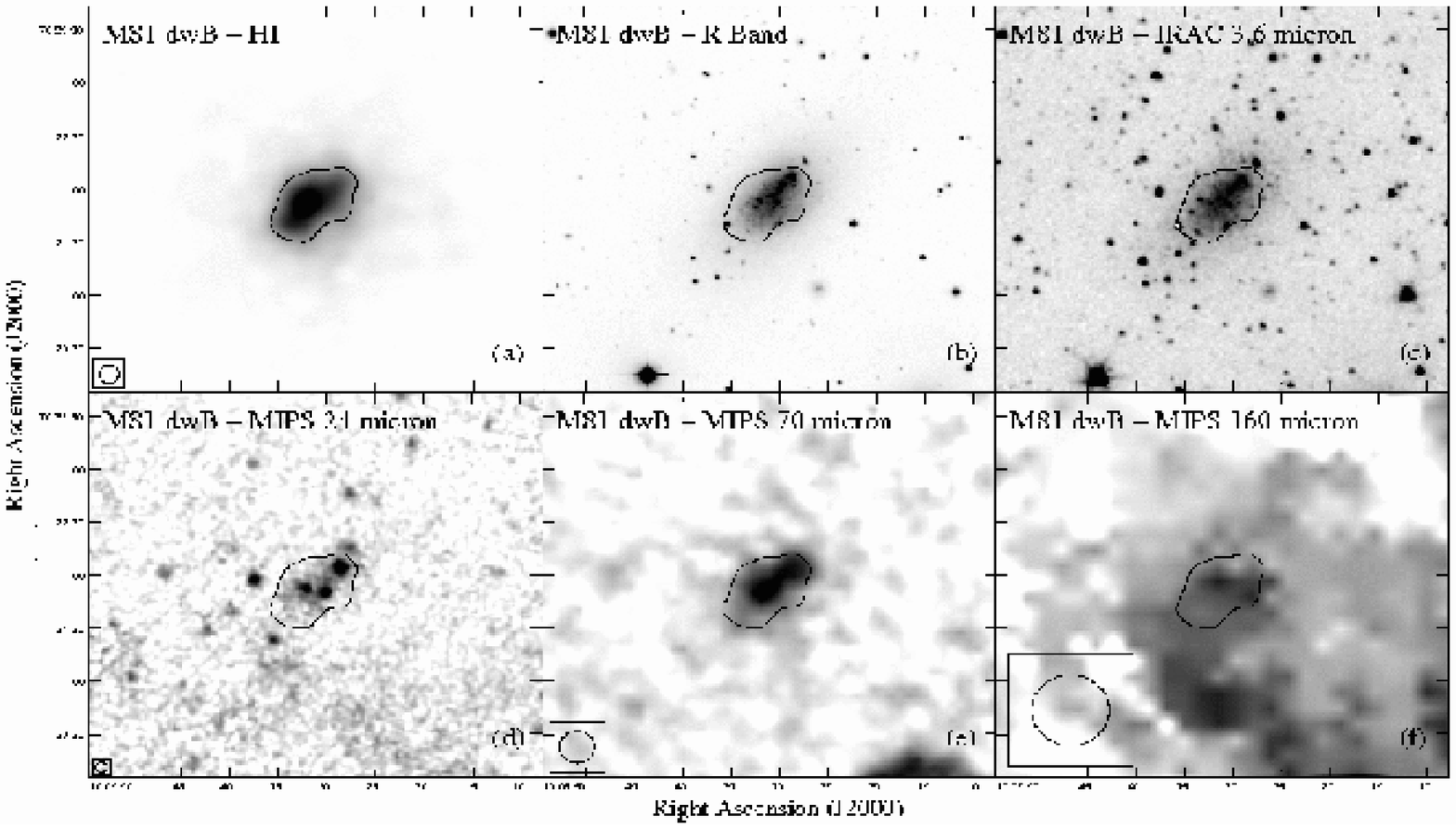}
\caption{M\,81\,dwB: (a) integrated {\em THINGS} \HI\ map (contour
  shown at N$_{\rm HI}$=$10^{21}\,$cm$^{-2}$ in all panels); (b)
  R--band image; (c) IRAC 3.6\mm\ image; (d), (e), and (f): 24\,\mm,
  70\,\mm\ and 160\,\mm\ images, respectively.  The sizes of the \HI\
  and MIPS beams are given in the lower left corners of their respective panels.}
\label{figcap6}
\end{figure*}

\begin{figure*}
\epsscale{0.8}
\plotone{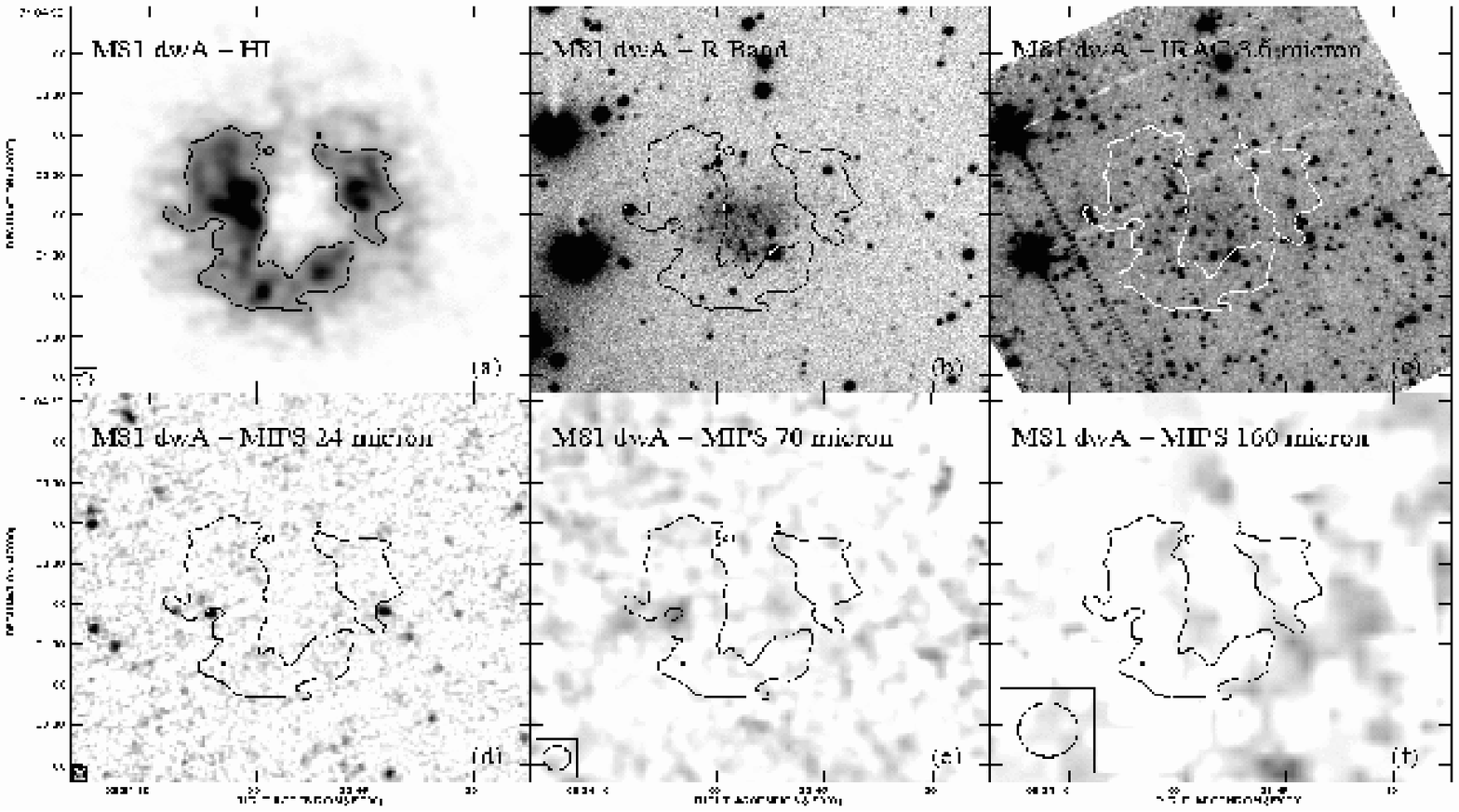}
\caption{M\,81\,dwA: (a) integrated {\em THINGS} \HI\ map (contour
  shown at N$_{\rm HI}$=3$\times10^{20}\,$cm$^{-2}$ in all panels);
  (b) R--band image; (c) IRAC 3.6\mm\ image; (d), (e), and (f):
  24\,\mm, 70\,\mm\ and 160\,\mm\ images, respectively. The sizes of
  the \HI\ and MIPS beams are given in the lower left corners of their
  respective panels. }
\label{figcap7}
\end{figure*}

\section{Dust and \HI\ Characteristics}
\label{S3}
\label{S3.1}

In Figs.~1--7 we present images of the individual galaxies (in order
of decreasing \HI\ mass). For each galaxy, we show six panels: (a) is
the integrated {\em THINGS} \HI\ map at 10$"$ resolution (only
exception: DDO\,165 beamsize: 21$''$); unless otherwise stated, one
\HI\ contour is drawn at N$_{\rm HI}$=1$\times10^{21}$\,cm$^{-2}$
(i.e., close to the canonical star formation threshold, e.g., Skillman
1996\nocite{skillman96conf}). An optical R--band and the {\it Spitzer}
IRAC band 1 (3.6\,\mm) image of the galaxies are shown in panels (b)
and (c) (the optical images have been observed either at the Calar
Alto 2.2\,m telescope or are taken from the ancillary {\it SINGS} data
archive).  Panels (d), (e) and (f) are the MIPS 24, 70 and 160\,\mm\
images of the same area. All panels show the same \HI\ contour as
presented in the first panel. The beamsizes for both the \HI\ and MIPS
images are shown in the lower left of the respective images.

\begin{figure}
\epsscale{0.9}
\plotone{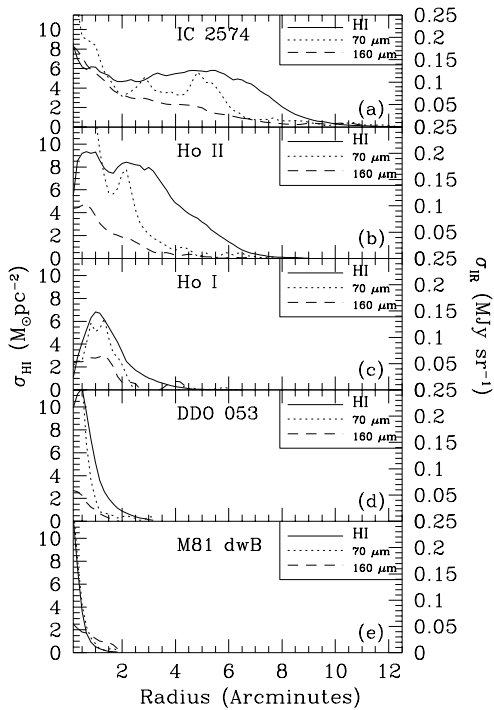}
\caption{Comparison of the \HI\ (solid), 70\,\mm\ (short dashed)
  and 160\,\mm\ (long dashed) radial profile of five M\,81 group dwarf
  irregular galaxies. The following deprojection parameters (position
  angle, PA; inclination $i$) where used to create these profiles:
  IC~2574 [PA=+60$^\circ$, $i$=57$^\circ$], Holmberg~II
  [PA=-15$^\circ$, $i$=26$^\circ$], Holmberg~I [PA=55$^\circ$,
  $i$=46$^\circ$], DDO\,53 [PA=--45$^\circ$, $i$=40$^\circ$],
  M\,81\,dwB [PA=--60$^\circ$, $i$=46$^\circ$].}
\label{figcap8}
\end{figure}

For each galaxy with {\it THINGS} \HI\ imaging, we compare the radial
profiles of the \HI, 70\,\mm\ and 160\,\mm\ MIPS images in
Fig.~\ref{figcap8} (as discussed for the individual systems below).
The deprojection parameters (inclination, position angle) used for the
creation of the radial profiles have been derived from the \HI\ maps
(see caption Fig.~\ref{figcap8}).

In the following, we briefly discuss the individual galaxies:

{\em IC~2574 (Fig.~\ref{figcap1}):} IC~2574 is the largest galaxy in
our sample and its \HI\ morphology is dominated by the presence of
\HI\ holes (Walter \& Brinks 1999).  The brightest region in the MIPS
bands is the supergiant shell (SGS) region in the north-east (Walter
\etal\ 1998); a spatially resolved {\it Spitzer} case study of this
SGS region is presented in Cannon et al.\ 2005\nocite{cannon05}. The
elevated emission toward the south--east corner in the 160\,\mm\ image
is caused by Galactic Cirrus emission (but this emission can be
separated from the emission of IC~2574).  Dust emission traced by the
70\,\mm\ emission is detected out to galactocentric radii of 7$'$
($\sim$7\,kpc, see Fig.~8).

{\em Holmberg~II (Fig.~\ref{figcap2}):} The distribution of \HI\ in
Holmberg~II is also characterized by the presence of numerous \HI\
holes \citep{puche92}. As in the case of IC~2574, the changing
morphologies in the individual MIPS bands and the corresponding
changing spectral energy distributions stress the importance of local
effects in characterizing the far--infrared emission (e.g., {Cannon
  \etal\ 2005}\nocite{cannon05}).  The radial surface brightness
profiles (Fig.~\ref{figcap8}) show that dust is detected out to at
least 4$'$ ($\sim$4\,kpc).

{\em Holmberg~I (Fig.~\ref{figcap3}):} The \HI\ distribution in
Holmberg~I is characterized by one giant \HI\ hole \citep{ott01}.
The \HI\ structure encompasses the optical emission and faint star
formation is present on the rim toward the south--east (see also
Sec.~\ref{S3.4}). This is the region where faint emission is
detected in all three MIPS bands.  At 70\,\mm, there is also diffuse
emission present toward the western \HI\ rim (though this
emission is of very low S/N).

{\em DDO\,165 (Fig.~\ref{figcap4}):} DDO\,165 shows extended emission
in both the \HI\ and the optical, but is not detected in the MIPS
bands. The brightest emission seen in the 70\mm\ and 160\mm\ bands
(toward the north) is a background galaxy (SDSS~J130639.44+674456.4 at
z=0.139). There is some emission seen at 24\,\mm\ and 70\,\mm\ toward
the centre of DDO~165 (not coincident with the peak of the \HI\
emission, and outside the main optical body of DDO\,165) -- future,
higher sensitivity observations (MIPS and \HI) are needed to see if
this emission is indeed physically related to DDO\,165.  Note that
DDO\,165 has one of the lowest star formation rates in our sample.

{\em DDO\,53 (Fig.~\ref{figcap5}):} The \HI\ distribution shows two
peaks, and the galaxy is detected in all three MIPS bands: The
brightest emission seen at 24\,\mm\ is associated with the northern
and southern \HI\ peak; the northern region is also the strongest in
the 70\,\mm\ image.  The 160\,\mm\ data are noisy, but 160\,\mm\
emission is still present in the regions seen in the 70\,\mm\ image.
The compact nature of DDO\,53 is also evidenced by the radial surface
brightness profiles shown in Fig.~8.

{\em M\,81\,dwB (Fig.~\ref{figcap6}):} M\,81\,dwB is the galaxy with
the lowest measured star formation rate in our sample.  The galaxy
shows a compact structure in \HI\ and the MIPS bands (cf.
Fig~\ref{figcap8}).  The detection at 160\,\mm\ is marginal (see
Tab.~1) and is surrounded by elevated background emission present.

{\em M\,81\,dwA (Fig.~\ref{figcap7}):} M\,81\,dwA is the faintest
dwarf in our sample.  Similar to Holmberg~I, the \HI\ distribution is
characterized by one large \HI\ shell which encompasses most of the
optical galaxy. No ongoing star formation has been detected in
M\,81\,dwA \citep{miller94}; it is also a MIPS non--detection.  This
may be explained by the fact that \HI\ column densities do not reach
values higher than N$_{\rm HI}$=5$\times10^{20}$\,cm$^{-2}$ in this
galaxy (i.e. the \HI\ column densities are below the canonical threshold
for star formation).

\subsection{HI Threshold for warm dust}
\label{S3.2}

\begin{figure}
\epsscale{0.9} 
\plotone{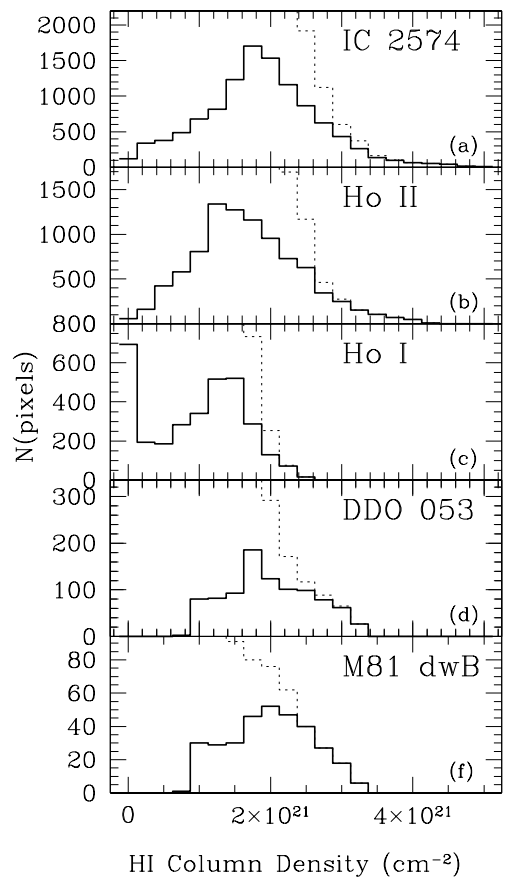}
\caption{Solid histogram: distribution of the 70\,\mm\ flux density
  above a threshold of 1.8 MJy\,sr$^{-1}$ ($\sim$5$\sigma$) as a
  function of \HI\ column density. Dashed histogram: total
  distribution of \HI\ column densities.}
\label{figcap9}
\end{figure}

From an inspection of the morphologies (Figs.~1--7) and radial
profiles (Fig.~\ref{figcap8}) it is clear that the dust emission
appears to be related to the distribution of the \HI, at least to
first order. To quantify this, we investigate if there is a certain
\HI\ threshold above which most of the dust emission is present. In
the following, we use the 70\,\mm\ data as a tracer for the warm dust
emission as they are (unlike the 24\,\mm\ data) not affected by the
presence of contaminating point sources (stars and background objects)
in the field.  Furthermore, they have higher resolution and
signal--to--noise than the 160\,\mm\ measurements.  In
Figure~\ref{figcap9}, we plot histograms of the distribution of the 70
\mm\ flux density above a threshold of 1.8~MJy\,sr$^{-1}$ (about the
5$\sigma$ level, i.e., encompassing most of the detected dust
emission) as a function of \HI\ column density for each galaxy (solid
histogram).  For each galaxy we also show the pixel--by--pixel
distribution of all \HI\ column densities (dashed histogram).  From
Fig.~9 (and the previous discussion on the relative distribution of
\HI\ and 70\,\mm\ emission) we draw the following conclusions:

-- Most of the detected 70\,\mm\ emission is coincident with \HI\
column densities of N$_{\rm HI}>10^{21}\,$cm$^{-2}$ (with a peak
around $1-2\times10^{21}\,$cm$^{-2}$). In the case of Holmberg~I, the
peak in the first histogram bin at low surface densities is spurious
and a result of the lower signal--to--noise in these data.

-- At high column densities ($>2.5\times10^{21}\,$cm$^{-2}$), the
solid (70\,\mm) and dashed (\HI) histograms follow each other closely,
implying that most of the high \HI\ column density regions are
associated with dust emission. In other words, there appear to be only
few regions of high \HI\ column densities that are {\em not}
associated with warm dust emission.

\begin{figure*}
\epsscale{0.8}
\plotone{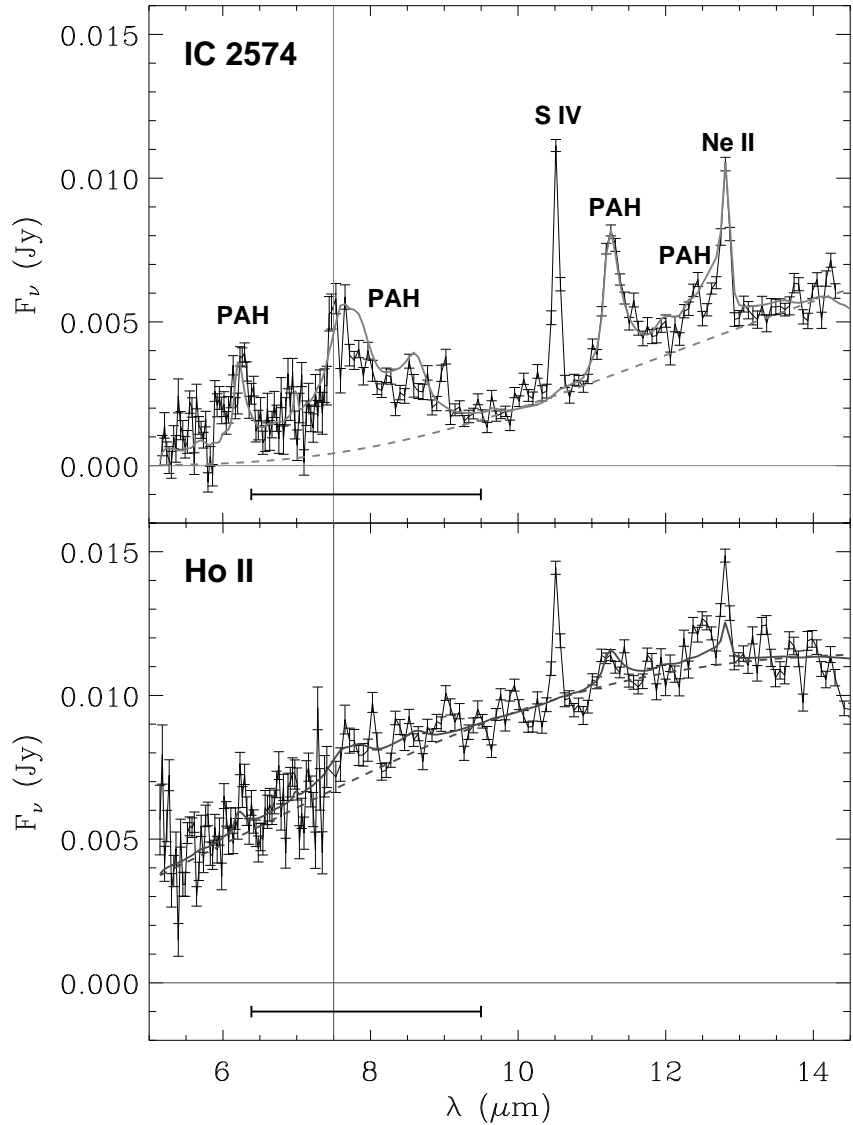}
\caption{IRS spectra of IC~2574 (top panel) and Holmberg~II (bottom
  panel). The vertical dashed line indicates the border of SL1 and SL2
  modules of the IRS spectrograph. The thick grey lines represent the
  sum of our scaled PAH template (derived from SINGS observations of
  NGC\,7552) and the underlying continuum (dashed curve, see text for
  details). The black horizontal line indicates the width ($>$10\%
  response) of the IRAC band 4 filter.}
\label{figcap10}
\end{figure*}

\subsection{IRS spectroscopy}

Selected regions in some of the galaxies in this sample have also been
observed with the Spitzer IRS spectrograph (Houck et al.\ 2004) as
part of SINGS. In the following we will discuss IRS 5--14\mm, low
resolution spectra obtained for specific regions in the two most
active galaxies in our sample, IC~2574 and Holmberg~II (Fig.~10, both
SFR$\sim$0.1\,M$_\odot$\,yr$^{-1}$). The spectra have been extracted
over a circular aperture of a diameter of $\sim$14$"$ that was
centered on the brightest region seen at 8\,$\mu$m in both galaxies
(see the caption of Table~2 for coordinates).  For comparison, Table~3
also summarizes some of the IRAC flux densities obtained for the same
apertures.

\subsubsection{IRS spectrum of IC~2574}

The spectrum of IC~2574 (upper panel in Fig.~10) clearly shows the
broad emission features from polycyclic aromatic hydrocarbons (PAHs)
which are typically found in mid-infrared spectra of massive
star--forming galaxies (e.g. Telesco 1988, Lu et al.\ 2003, Smith et
al.\ 2006).  To derive a simple template for the aromatic/PAH
features, we have taken the IRS spectrum of the SINGS galaxy NGC\,7552
(12+log(O/H)$\sim$8.5, Moustakas et al.\ 2007,
SFR$\sim$7\,M$_\odot$\,yr$^{-1}$) which shows a typical spectrum with
one of the highest signal--to--noise ratios in the SINGS sample (Smith
et al.\ 2006). From this spectrum we subtracted a 200\,K blackbody
spectrum that has been normalized to the pseudo-continuum at 10\,\mm\
and 13.5\,\mm. This PAH template was then scaled to fit the PAH
feature at 11.3\,\mm\ in IC~2574 after a scaled blackbody spectrum of
200\,K (shown as the dashed thick line) has been added (thick grey
line, normalized in the same way as for NGC\,7552).  Although a
blackbody is an unphysical representation of the dust continuum
beneath the emission bands, this simple procedure allows us to compare
the strength of the PAH bands in different sources in a consistent
way. The relatively low signal--to--noise ratio of the spectrum
prevents a full spectral decomposition (e.g., as done by Smith et al.\
2006).

It is interesting to note that this simple template fits the spectrum
of IC~2574 quite well to first order. Even though the S/N ratio is
low, the 7.7\mm/11.3\mm\ PAH ratio in IC~2574 appears to be lower
than in NGC\,7552 -- band--variations like that are known for many
other galaxies (e.g., Draine \& Li 2001, Vermeij et al. 2002, Cannon
et al.\ 2006, see detailed discussion in Smith et al.\ 2006).
Engelbracht et al. (2005) showed that there appears to be a
metallicity threshold of 12+log(O/H)$\sim$8.2 below which PAH emission
is not detected in galaxies\footnote{Note that the oxygen abundances
  used by Engelbracht et al.\ (2005) were based on a heterogeneous
  compilation of measurements from the literature, whereas the
  abundances in our study have been derived self--consistently, and
  placed on a common abundance scale (see Moustakas et al.\ 2007 for
  details)}.  The gas--phase metallicity of IC~2574 (Table~1) is
slightly below their threshold -- the fact that we do see PAH emission
in IC~2574 is likely not due to metallicity variations within the
galaxy (as no evidence of significant local metallicity variations has
been found in dwarf galaxies; e.g., Kobulnicky \& Skillman 1996,
1997), but rather due to to other factors that influence the aromatic
feature emission (e.g., radiation field, geometry, elemental
composition of the ISM).

Two bright emission lines are present as well in the spectra of
IC~2574: [Ne~II] emission at 12.8\,\mm\ and [S~IV] emission at
10.5\,\mm.  The latter line is not present in the template spectrum of
NGC\,7552 and indicates the presence of massive stars -- this line is
usually found to be faint in massive galaxies (e.g., see discussion in
Rigby \& Rieke 2004), but bright in the highly ionized gas of blue
compact dwarfs (e.g., Madden et al.\ 2006).

\subsubsection{IRS spectrum of Holmberg~II}

The IRS spectrum of Holmberg~II (lower panel in Fig.~10) is markedly
different from that of IC~2574; the spectrum is dominated by
continuum emission, and only very faint PAH features are present. The
thick grey line again shows the comparison to our simple template:
Here two blackbody curves (one at 300\,K and one at 700\,K,
contributing equal flux density at 8.3\,$\mu$m) were needed to
adequately fit the continuum of Holmberg~II (note again that these
curves do not represent physically meaningful numbers but are only
used to describe the shape of the continuum to first order, see also
discussion in Smith et al.\ 2006); we then added a scaled version of
our PAH template to this continuum emission (thick grey line). A
comparison to this curve shows that, although the S/N is low, faint
PAH features appear to be present at 6.2, 7.7, 11.3 and 12.7\mm --
e.g., the 11.3\mm\ PAH feature is detected at 7$\sigma$ (total flux;
peak: 4$\sigma$). From this it is clear that the PAH--to--continuum
ratio in Holmberg~II is much lower than in the case of IC~2574 as
discussed below. We note that the gas--phase metallicity of
Holmberg~II (Tab.~1) is lower than in IC~2574 by nearly a factor of
two (below the threshold derived by Engelbracht et al., 2005).  This
result emphazises the fact that the strength of the PAH features is
not a simple linear function of metallicity (see also Smith et al.\
2006, who find a wide range of PAH strengths (factor of $\sim$ 10)
near the Engelbracht et al.\ threshold).  As in the case of IC~2574,
line emission from [Ne~II] and [S~IV] is detected in Holmberg~II.

\begin{deluxetable*}{lccccc}         
  \tabletypesize{\scriptsize} \tablecaption{IRS vs.\ IRAC flux
  densities} \tablewidth{0pt} \tablehead{ \colhead{Galaxy}
  &\colhead{F(8\mm)} &\colhead{F(6\mm)} &\colhead{F(3.5\mm)}
  &\colhead{F(8\mm)$_{\rm IRS}$/F(8\mm)$_{\rm IRAC}$}
  &\colhead{F(6\mm)$_{\rm IRS}$/F(6\mm)$_{\rm IRAC}$}\\ \colhead{}
  &\colhead{mJy} &\colhead{mJy} &\colhead{mJy} & \colhead{} &
  \colhead{}}
  \startdata
  IC~2574$^{\rm a}$ &   3.10  &   1.39  &   0.73  &   0.89    &   0.95\\
  Holmberg~II$^{\rm a}$ & 8.53 & 5.75 & 2.02 & 0.93 & 0.81 \enddata
  \tablenotetext{a}{Central coordinates (J2000.0) and diameters $D$ of
    circular apertures: IC~2574: RA: 10~28~48.3, DEC: +68~28~03,
    $D$=13.67$''$, Holmberg~II: RA: 08 19 12.8, DEC: +70 43 08,
    $D$=14.63$''$.}
\label{t2}
\end{deluxetable*} 

\subsubsection{PAH--to--Continuum Ratios}

Using our simple decomposition of the PAH features and the continuum
emission, we can now constrain the PAH--to--continuum ratios for both
galaxies. We do this for two bands: A) the IRAC band 4 (`8\,\mm\
band'), encompassing the broad PAH features at 7.7\mm\ and 8.6\mm\
(see horizontal line in Fig.~10 for the wavelength range covered by
the 10\% response of the IRAC band 4) and B) the PAH band at 11.3\mm\
(here integrated within a mock square filter between 10.8 and
11.8\mm). We also calculate the PAH--to--total IR luminosity ratios
for both regions below (but note that the method employed here is not
directly comparable to the one used in Smith et al.\ 2006).

{\em IC~2574}: From the IRS spectrum we derive a flux density of
2.7\,mJy for the IRAC band 4 bandpass which is in good agreement with
the value derived from the actual IRAC band 4 measurement (3.0\,mJy,
see Table~3).  The continuum contribution is 0.7\,mJy and the
contribution from the PAH feautures is 2.0\,mJy; i.e. we derive a
PAH--to--continuum ratio for this spectral region of $\sim$2.9. For
the 11\,\mm\ feature we derive a ratio of 0.5 (continuum: 3.3\,mJy,
PAH: 1.7\,mJy).  Using our definitions for the 8\mm\ and 11\mm\
bandpasses and the total infrared (TIR) luminosities in this aperture
(L$_{\rm TIR}$, derived from the MIPS images and using the relation in
Dale \& Helou 2002) we get the following ratios: L$_{\rm PAH, 8\mu
m}$/L$_{\rm TIR}$=0.0089, L$_{\rm PAH, 11\mu m}$/L$_{\rm TIR}$=0.0011.
  
{\em Holmberg~II}: From the IRS spectrum we derive a flux density of
7.9\,mJy for the IRAC band 4 bandpass which is in good agreement with
the value derived from the actual IRAC band 4 measurement (8.53\,mJy,
Table~3).  The continuum contribution is 7.2\,mJy, the contribution
from PAHs is 0.7\,mJy, leading to a PAH--to--continuum ratio of 0.1.
For the 11\,\mm\ feature the ratio is even lower, 0.02 (continuum:
10.4\,mJy, PAH: 0.2\,mJy). The corresponding ratios compared to the
total infrared luminosity in this aperture are: L$_{\rm PAH, 8\mu
  m}$/L$_{\rm TIR}$=0.0037, L$_{\rm PAH, 11\mu m}$/L$_{\rm
  TIR}$=0.0002. Given the faintness of the PAH features in
Holmberg~II, these values are uncertain (by $\sim$50\%).

For comparison, we also derive the PAH-to-continuum ratios for our
template galaxy NGC\,7552 and get flux density ratios of 6.3 and 0.9
for the IRAC4 and [10.8; 11.8] $\mu$m bandpasses, respectively. For
the TIR luminosities we get the following ratios for NGC\,7552:
L$_{\rm PAH, 8\mu m}$/L$_{\rm TIR}$=0.070, L$_{\rm PAH, 11\mu
  m}$/L$_{\rm TIR}$=0.0077; i.e. the ratio in IC~2574 is about a
factor of 7 (Holmberg~II: factor of $>$20) less than in our template
galaxy (see Smith et al.\ 2006 for variations within the SINGS
sample).

\begin{figure*}
\plottwo{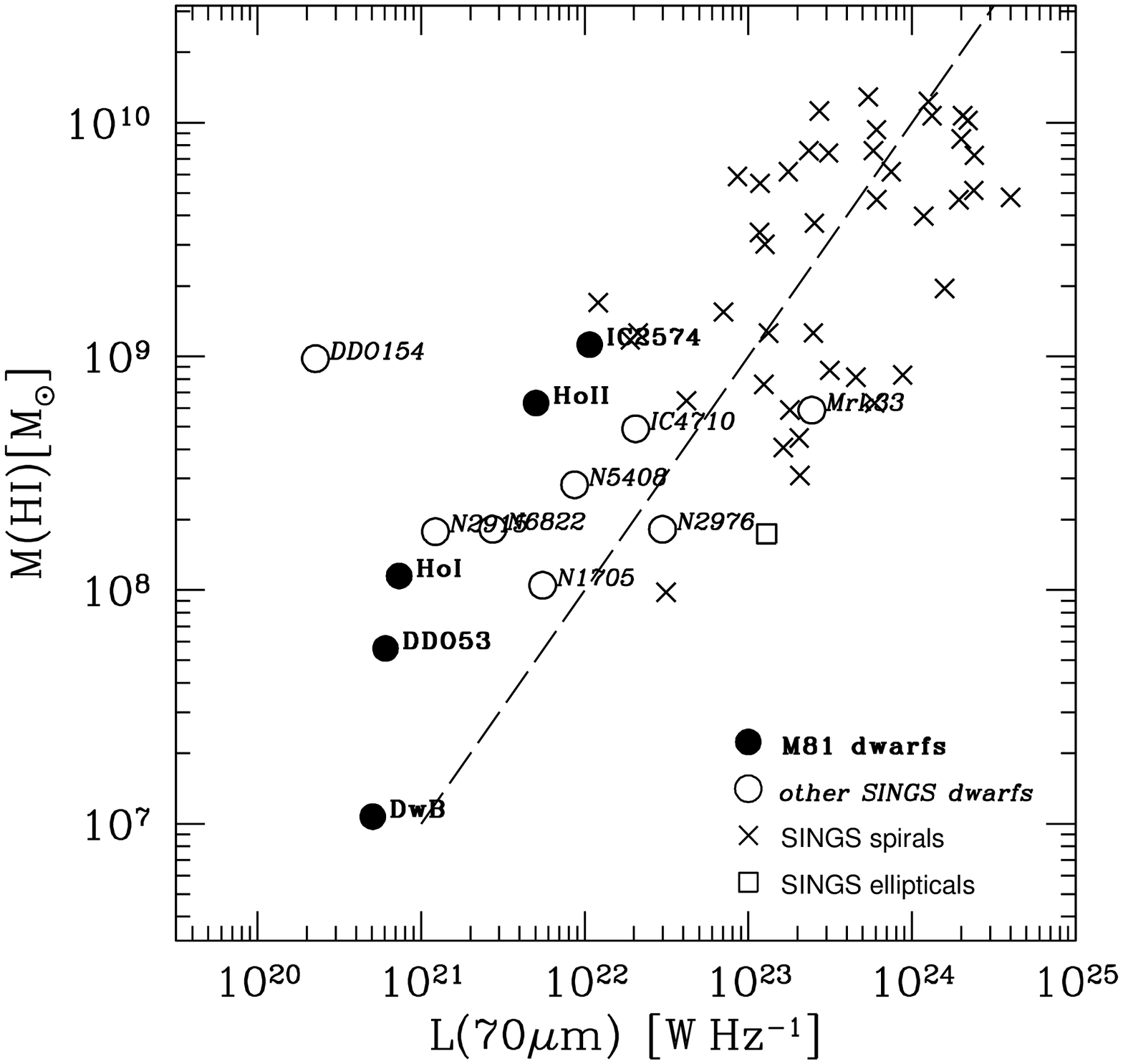}{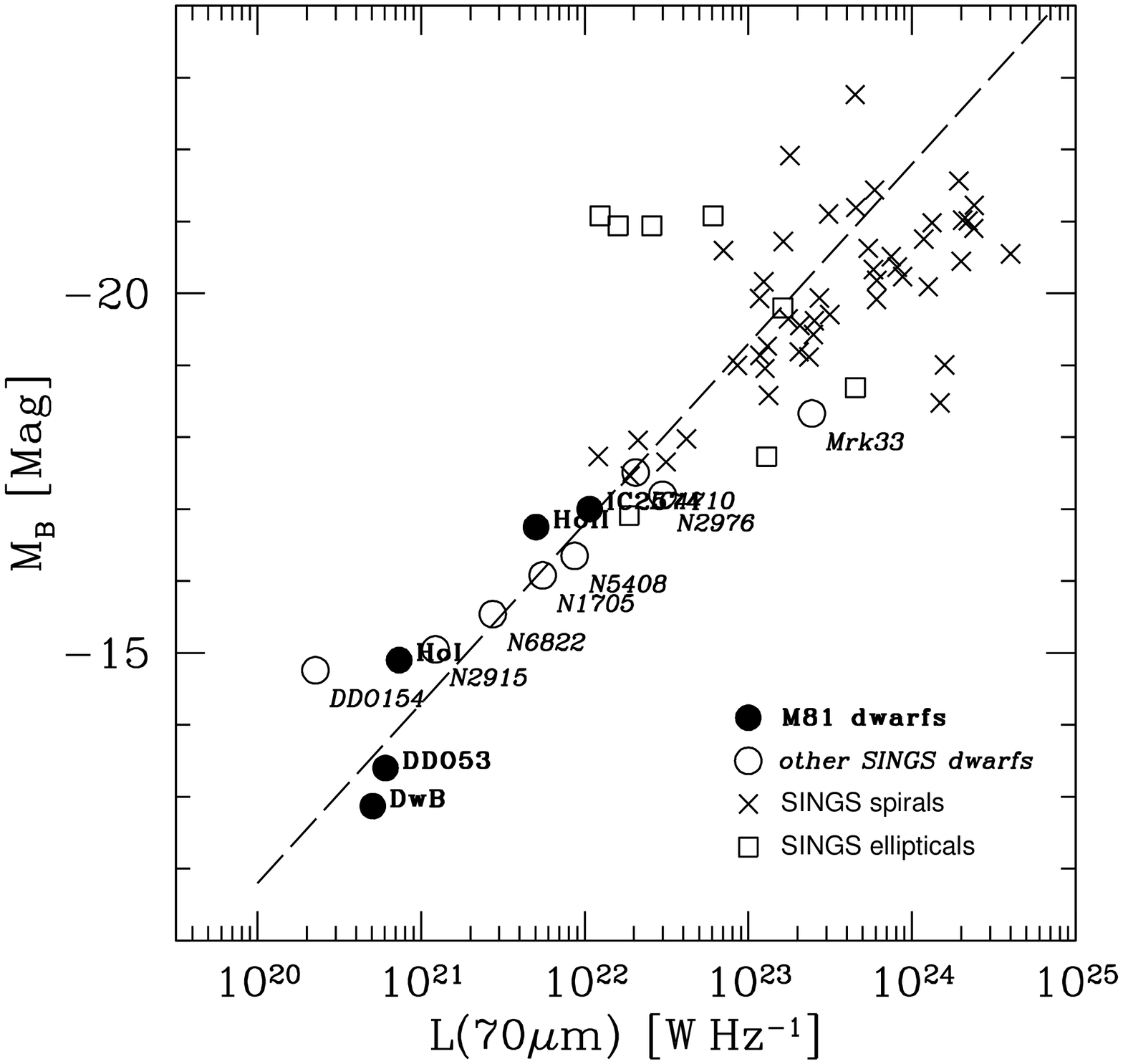}
\caption{(a): \HI\ mass as a function of 70\,\mm\ specific luminosity. (b): 
  Blue magnitudes (M$_{\rm B}$) as a function of 70\,\mm\ specific
  luminosity. The dashed line in both plots represents a linear
  relationship between the two variables.}
\label{figcap11}
\end{figure*}

\subsection{Global relations for the M\,81 group dwarfs}
\label{S3.3}

In the following we will compare the properties found for the M\,81
dwarf irregular galaxies to other galaxies in the {\it SINGS} sample
(Kennicutt et al.\ 2003). To do so, we have divided the {\it SINGS}
galaxies into four categories: the M\,81 group dwarf irregular
galaxies of this study (shown as filled circles in the following
plots), other dwarf galaxies (open circles), elliptical/S0's (open
squares) and spiral galaxies (crosses). All MIPS flux densities for
the {\it SINGS} galaxies are taken from Dale et al.\ (2006), and the
gas--phase metallicities are taken from Moustakas et al.\ (2007, based
on the Pilyugin \& Thuan 2005 strong--line abundance calibration).

As there apparently exists some correlation between the \HI\ and the 70
\mm\ emission (see discussion in the previous sections), we start by
comparing the \HI\ masses to the 70\,\mm\ specific
luminosity\footnote{In the following we plot the specific luminosities
(units of W\,Hz$^{-1}$). To derive the luminosities in a given MIPS
band, this number needs to be multiplied by the effective bandwidth of
the MIPS filter (in Hz). Note that some authors (e.g.  {Calzetti
\etal\ 2005}\nocite{calzetti05}) define the luminosity as the specific
luminosity times the observed frequency (in Hz).}  in
Figure~\ref{figcap11}a. It is obvious that there is a large scatter
between the two quantities (the dashed line indicates a linear
relationship). If we plot the absolute blue magnitudes (M$_{\rm B}$,
taken from Moustakas et al.\ 2007) as a function of the 70\,\mm\
specific luminosity instead (Figure~\ref{figcap11}b), this relation
gets tighter (dashed line).  The larger scatter in the \HI--70\mm\
relation is due to the fact that the dwarf irregular galaxies of our
sample have more \HI\ mass per blue magnitude compared to more massive
spirals.  This fact, i.e.  that M$_{\rm HI}$/M$_{\rm B}$ increases for
dwarf irregular galaxies was noted long before (e.g., Skillman
1996\nocite{skillman96conf}).

\begin{figure*}
\plottwo{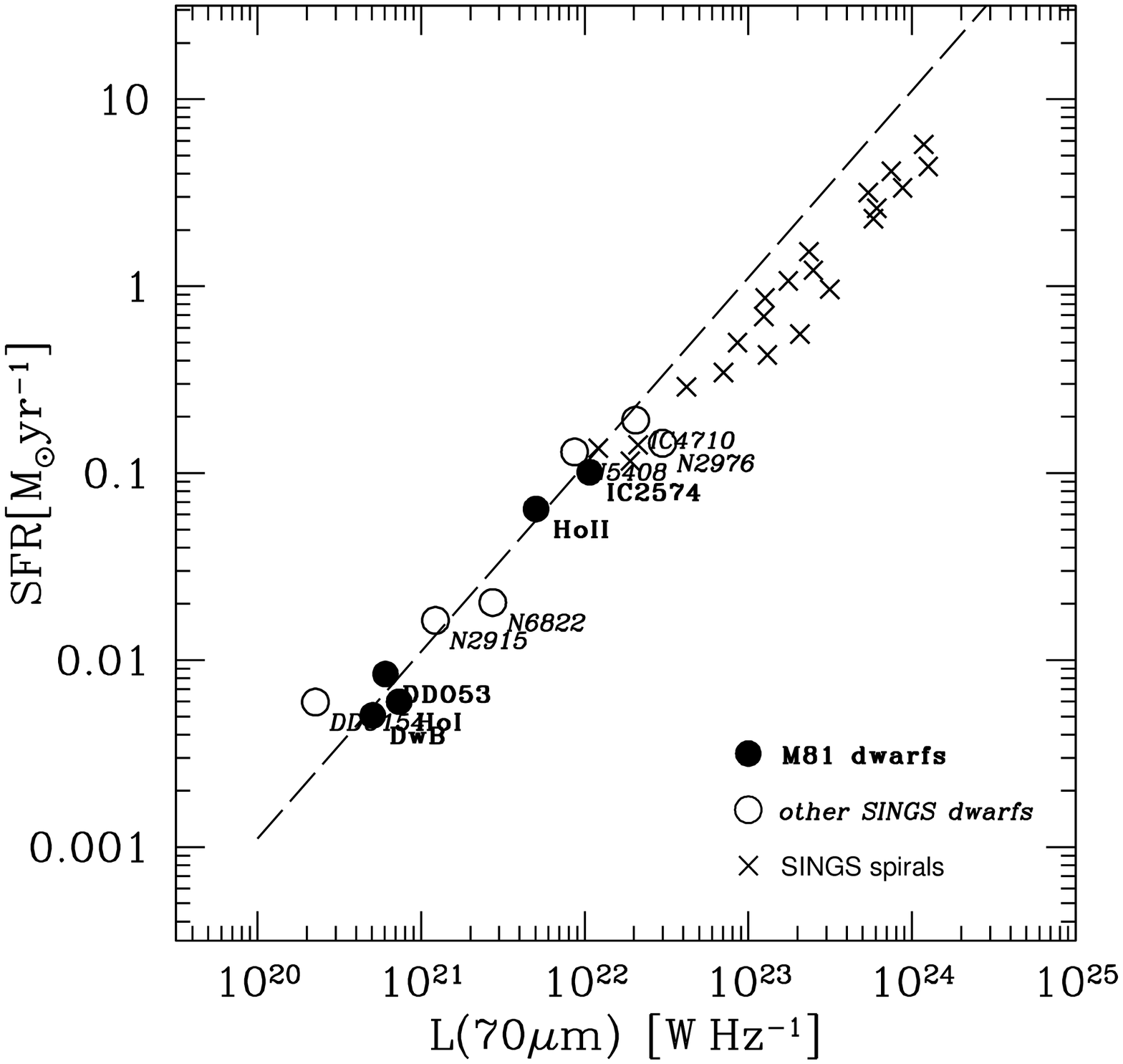}{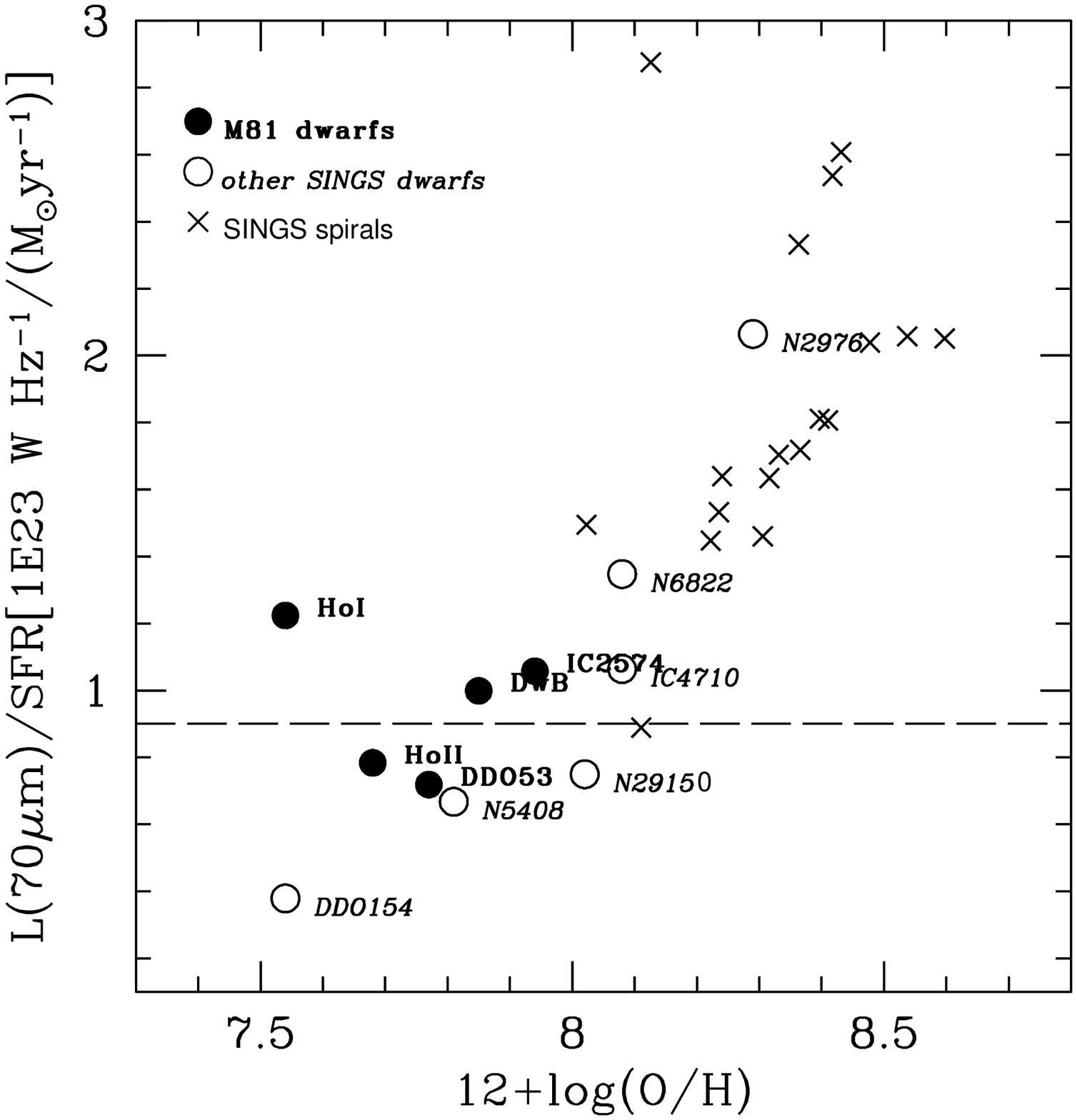}
\caption{(a): SFR as a function of 70\,\mm\ specific luminosity. 
  (b): Ratio of 70\,\mm\ specific luminosity and SFR as a function of
  metallicity.  The dashed line in the left panel represents a linear
  relationship between the two variables (same relation as shown in the
  right hand panel).}
\label{figcap12}
\end{figure*}

As a next step, we plot the star formation rates (SFRs) of individual
galaxies as a function of the 70\,\mm\ specific luminosity
(Fig.~\ref{figcap12}a).  The SFRs were derived using SFR =
L(H$\alpha_{\rm corr}$)/1.26$\times10^{41}$\,erg\,s$^{-1}$ [M$_\odot$
yr$^{-1}$] \citep{kennicutt98} (assuming solar metallicity and the
Salpeter (1955) initial mass function between 0.1 and 100\msun).
L(H$\alpha_{\rm corr}$) is the H$\alpha$ luminosity corrected for
extinction within the galaxy.  Here we use the relation derived for
SINGS galaxies by Kennicutt et al.\ (2006) and Calzetti et al.\
(2006): F(H$\alpha$)$_{\rm corr}$=F(H$\alpha$)$_{\rm
  obs}$+0.035$\times$F'(24\mm).  F(H$\alpha$)$_{\rm obs}$ is the
observed H$\alpha$ flux (in erg\,s$^{-1}$\,cm$^{-2}$, corrected for
Galactic absorption and contribution from [N II]) taken from Kennicutt
et al.\ 2006 and Lee 2006; F'(24\mm) is defined as
F'(24\mm)=1.25$\times$10$^{13}$Hz$\times$ F(24\mm)[Jy] (i.e. F(24\mm)
multiplied by the observed frequency in Hz).  SFRs were only derived
for those SINGS galaxies for which accurate H$\alpha$ measurements are
available. The tighter relation between L(70\mm) and SFR (as compared
to the M$_{\rm B}$-- L(70\mm) relation discussed above), can (to first
oder) be explained by the fact that, typically, 70\mm\ is close to the
peak of the FIR emission and thus is a measure of the total bolometric
luminosity which is expected to scale with the SFR.

However, from Figure~\ref{figcap12}(a) it is also obvious that the
galaxies do not follow a linear relation over all luminosities (as
indicated by the dashed line). To exemplify this, the ratio of the
70\,\mm\ specific luminosity and the SFR is plotted in
Figure~\ref{figcap12}(b) as a function of the gas--phase oxygen
abundance (taken from Moustakas et al.\ 2007). If there was a linear
relation between the 70\,\mm\ emission and the SFR, one would expect
the symbols to lie near a horizontal line in this plot.  However, the
dwarfs, shown as circles, clearly occupy a lower L$_{70\mu m}$/SFR
space compared to the more massive galaxies. On average, the dwarf
galaxies appear to be underluminous in 70\,\mm\ emission relative to
their SFR by a factor of $\sim2$. A similar conclusion (dwarfs have
lower L$_{\rm IR}$/L$_{\rm H\alpha}$) has been reached by {Hunter
\etal\ 1989}\nocite{hunter89}, albeit for more luminous systems.  This
behavior may, to first order, be attributed to the fact that these
objects have low metallicities and low dust contents (see discussion
in Sec.~3.5 and Draine et al.\ 2007). However it is also clear that
the scatter is large and that there is no simple relation between
oxygen abundance and 70\mm\ luminosities per unit SF.

\begin{figure*}
\plottwo{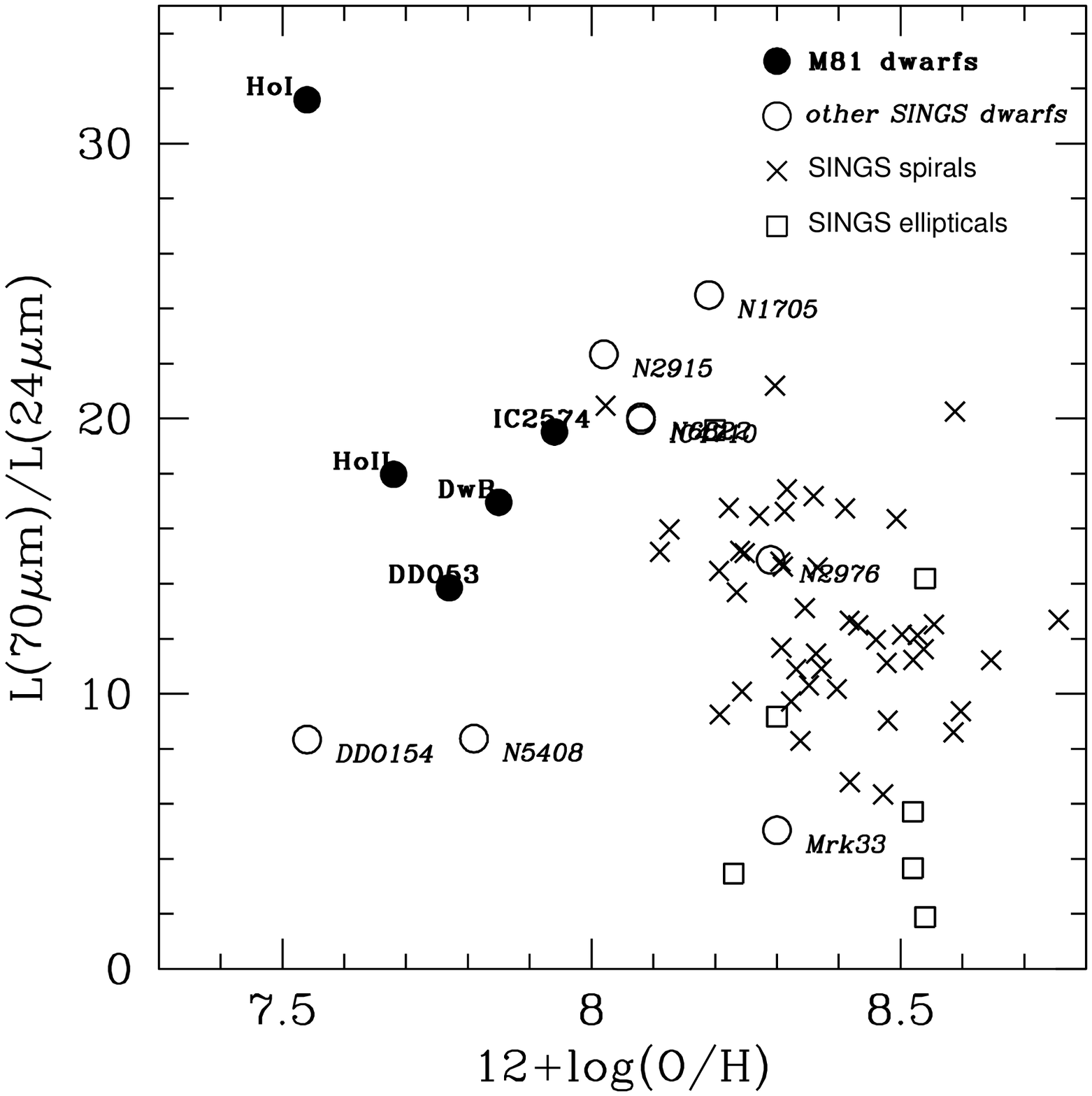}{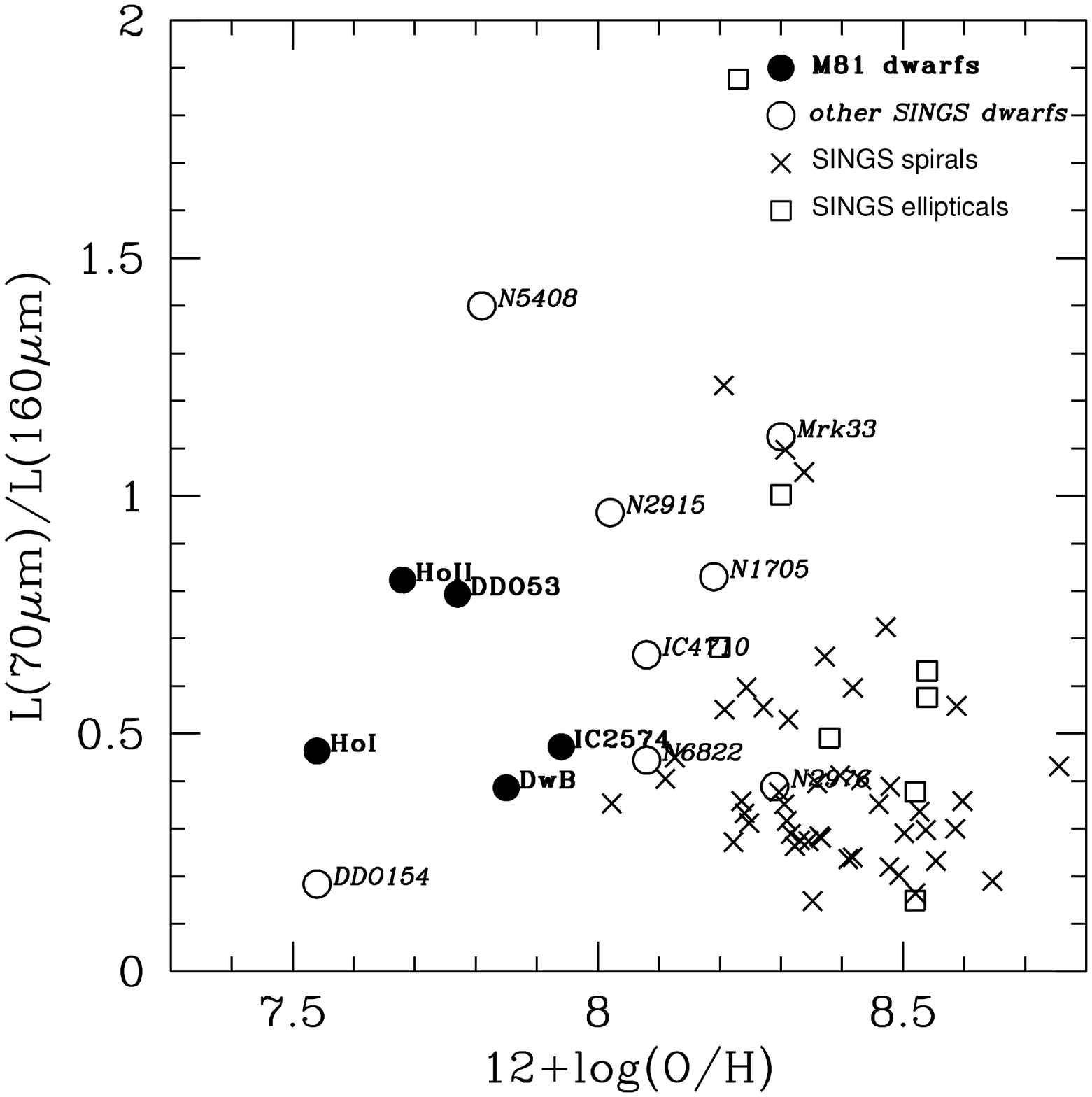}
\caption{(a): 70\,\mm\ to 24\,\mm\ flux density ratio as a function of oxygen
  abundance.  (b): 70\,\mm\ to 160\,\mm\ flux density ratio as a
  function of oxygen abundance.}
\label{figcap13}
\end{figure*}

We now investigate how the global MIPS colors (here: the 70\,\mm/160
\mm\ and 70\,\mm/24\,\mm\ ratios) of the M\,81 groups dwarfs compare
to the other {\it SINGS} galaxies.  In Figure~\ref{figcap13} we plot
both ratios as a function of the oxygen abundance: Although the
scatter is large, the dwarf galaxies have elevated 70\,\mm/160\,\mm\
ratios and 70\,\mm/24 \mm\ ratios compared to the spiral galaxies in
{\em SINGS}.  The elevated 70\,\mm/160\,\mm\ ratios imply that the
effective temperature of the dust in the dwarf irregular galaxies is
on average higher than in more massive spirals.  In this simplistic
picture, the higher effective temperature results in a peak of the SED
that is shifted toward the 70\,\mm\ waveband (cf. {Hunter \etal\
1989}\nocite{hunter89}, {Dale \etal\ 2005}\nocite{dale05}). In this
context it is interesting to keep in mind that the L$_{70\mu m}$/SFR
ratio in dwarfs is lower than in the spirals (see above). In other
words, if the dust temperature in our sample dwarfs were the same as
in the spirals, the L$_{70\mu m}$/SFR ratio would decrease even
further.

The elevated 70\,\mm/24\,\mm\ ratio in our sample dwarfs is more
difficult to interpret as the origin of the 24\mm\ emission is not
certain. According to the models by Draine \& Li (2006), this emission
is due to in part to single--photon heating, although in galaxies with
strong 24\mm\ emission it is primarily due to warm grains in strong
radiation fields. The 24\mm\ luminosity thus depends on the intensity
of the radiation heating the dust (which depends on the density in the
HII regions, the degree of clustering of O stars, as well as on the
dust abundance).  Future detailed modelling of the SEDs is needed to
fully describe this behavior. We also note that our results do not
necessarily hold for all classes of dwarf galaxies; i.e., extreme
cases such as the metal--poor blue compact dwarf SBS~0335-052 have
more extreme colors (very low 70\,\mm/24\,\mm\ ratio, Houck et al.\
2004).

\subsection{Spatial comparison to H$\alpha$}
\label{S3.4}

\begin{figure*}
\epsscale{1.0}
\plotone{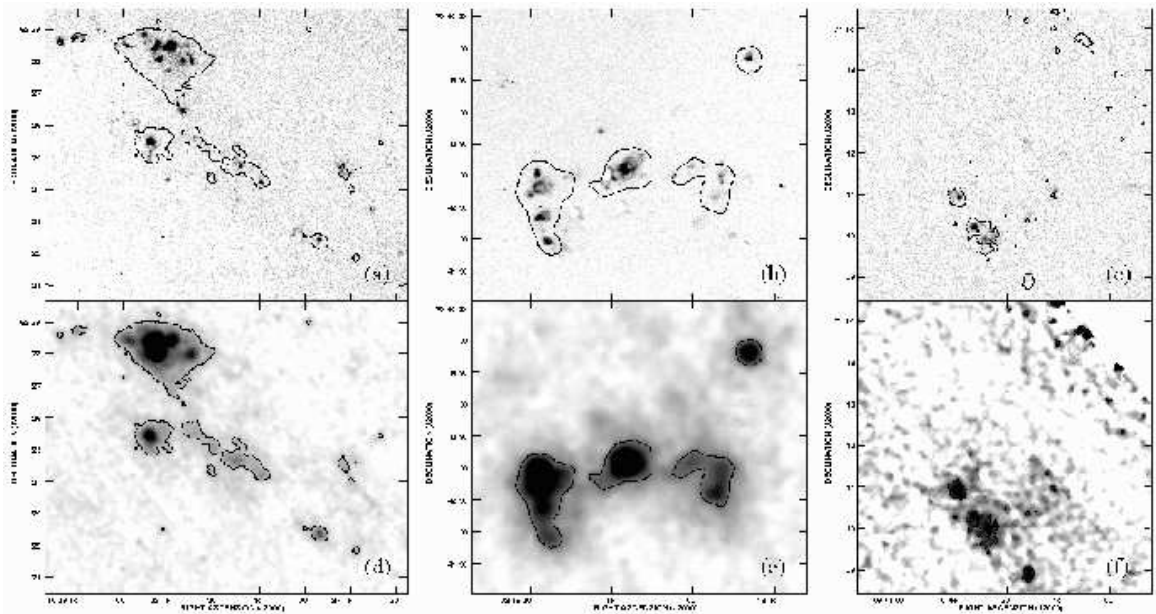}
\caption{{\em Top:} H$\alpha$ images for the three brightest dwarf
  galaxies in our sample (a:~IC~2574; b:~Holmberg~II,
  c:~Holmberg~I).  Contours indicate 70\,\mm\ surface brightness
  levels [(a): 1.8\,MJy\,sr$^{-1}$, (b): 4\,MJy\,sr$^{-1}$, (c):
  2.6\,MJy\,sr$^{-1}$. {\em Bottom:} 70\,\mm\ maps of the three
  galaxies with the same contour overlaid (the emission seen toward
  the north--west of Holmberg~I is noise).}
\label{figcap14}
\end{figure*}

The apparent strong correlation between the 70\,\mm\ luminosities and
the SFR implies that the ongoing star formation is the main heating
source for the warm dust.  We now compare the spatial distribution of
the H$\alpha$ and the 70\,\mm\ emission: Figure~\ref{figcap14} (top)
shows the H$\alpha$ images for the three brightest H$\alpha$ emitters
in our sample (IC~2574, Holmberg~II and Holmberg~I).  The contour
shown in the invidual panels represents low--level 70 \mm\ emission
(shown in Figure~\ref{figcap14}, bottom).  From this we find a very
good correlation between the locations of strong H$\alpha$ and
70\,\mm\ emission. This has been found in other galaxies before (e.g.,
M\,51: {Calzetti \etal\ 2005}\nocite{calzetti05}) but it is
interesting to note that the same holds true even for faint dwarfs
such as Holmberg~I.  We note that there is also diffuse dust emission
present in our objects (e.g., Holmberg~II, IC~2574) that is likely
heated by the underlying stellar population or by UV photons that are
leaking from the HII regions (see {Cannon \etal\
2006b}\nocite{cannon06b} for a detailed discussion on the diffuse dust
component in the Local Group galaxy NGC\,6822, see also, e.g., Popescu
et al.\ 2002, Popescu \& Tuffs 2003, Hinz et al.\ 2006).

\begin{deluxetable*}{lccccc}
  \tabletypesize{\scriptsize} \tablecaption{Dust and Gas Masses} \tablewidth{0pt} \tablehead{
    \colhead{Galaxy\tablenotemark{a}} 
    &\colhead{M$_{\rm dust}$\tablenotemark{b}}
    &\colhead{M$_{\rm HI}$}
    &\colhead{M$_{\rm HI}^{\rm apt, c}$}
    &\colhead{M$_{\rm dust}$/M$_{\rm HI}$}  
    &\colhead{M$_{\rm dust}$/M$^{\rm apt}_{\rm HI}$} \\
    \colhead{} 
    &\colhead{(10$^4$ M$_{\odot}$)}
    &\colhead{(10$^8$ M$_{\odot}$)}    
    &\colhead{(10$^8$ M$_{\odot}$)}
    &\colhead{10$^{-3}$} 
    &\colhead{10$^{-3}$}}
\startdata
  IC~2574        &72   & 14.75 & 2.8 & 0.49 & 2.6\\
  Holmberg~II     &12   &  5.95 & 1.2 & 0.20 & 1.0\\
  Holmberg~I      &6.8  &  1.40 & 0.4 & 0.48 & 1.7\\
  DDO053          &1.0  &  0.60 & 0.2 & 0.17 & 0.5
 \enddata
\label{t3}
\tablenotetext{a}{See Table 1 for galaxy parameters}
\tablenotetext{b}{Dust mass derived using the SED models of Draine \& Li 2006}
\tablenotetext{c}{HI mass in region where dust emission is present}
\end{deluxetable*}

\subsection{Dust Masses}
\label{S3.5}

We now discuss the dust masses of the M\,81 group dwarf irregular
galaxies. As the dust mass derivations mostly rely on the longest
wavelength MIPS bands (i.e. the 70\,\mm\ and 160\,\mm\ bands), and
given the large error bars in the measurements of the dwarf galaxies
discussed here (in particular in the 160\,\mm\ band, see
Table~\ref{t1}), the dust masses presented in the following can only
be considered to be order of magnitude estimates. Also, the
observations are not sensitive to cold dust, which emits at
wavelengths $>>$160\,\mm\ (i.e., there may be more dust present that
is not heated by star formation).

To estimate the dust masses we use the new model presented in Draine
\& Li that include variable PAH abundances (2006, which is based on
the models presented in {Li \& Draine (2001,
  2002)}\nocite{li01,li02}). In their model, radiation field strengths
are varied via power-law distributions; PAH, silicate and graphite
grains are illuminated and the resulting SEDs can be compared to the
observations. Table~3 (column 2) summarizes the dust masses of the
sample dwarf galaxies using their models (note that DDO\,165 and
M\,81\,dwA were non--detections; the flux density measurements of
M\,81\,dwB are too uncertain to derive a meaningful dust mass): the
dust masses span almost two orders of magnitude from
$\sim$1.0--70$\times10^4$ \msun. If we use the SED models by
\citet{dale01} and \citet{dale02} we get masses that are higher by a
factor of $\sim$5.  We attribute the difference to the fact that their
models and dust mass correction factors were developed for normal
galaxies and are therefore likely not appropriate for dwarfs.

With these dust masses we can now estimate the dust--to--gas ratios of
the dwarf galaxies in our sample. In the following we define M$_{\rm
gas}$=M$_{\rm HI}$ for the dwarfs, i.e. we do not take the possible
(unknown) contribution of molecular gas into account (and we do not
correct for the contribution of Helium).  Using our dust and \HI\
masses (Table~3, column 3), we derive an average M$_{\rm
dust}$/M$_{\rm HI}$ ratio of $\sim3\times10^{-4}$ (numbers for the
individual galaxies are given in Table~3, column 4). These values do
not appear to be a function of the measured metallicities (however the
spread in metallicies in the remaining sample is small).  If we
consider only the \HI\ mass of the M\,81 group dwarfs in an aperture
defined by the extent of the dust emission (M$_{\rm HI}^{\rm apt}$,
Table~3, column 5) this ratio increases to an average of M$_{\rm
dust}$/M$_{\rm HI}^{\rm apt}\sim1.5\times10^{-3}$ (Table~3, column 6).
This is because, on average, about three quarters of the extended
\HI\ mass has no associated bright dust emission (cf.\ Figs.~1--7).

Draine \etal\ (2007, in prep.) present detailed dust mass estimates
for all {\it SINGS} galaxies and derive a typical spread in
dust--to--gas ratios for the more massive galaxies of 0.005$<$M$_{\rm
dust}$/M$_{\rm gas}<$0.02 (here including the contribution of
molecular hydrogen). The value for the dwarfs is thus about an order
of magnitude lower; it would decrease further if the dwarfs had a
significant component of molecular hydrogen. The ratio for the dwarfs
would increase, however, if they had an additional cold dust component
(e.g., in the outskirts where the stellar radiation field is low, see
also Draine et al.\ 2007) that can not be traced with the available
MIPS observations. E.g., in the case of NGC\,1569, Galliano et al.\
2003 report a millimetre excess in the dust SED -- this cold dust
component could account for up to 70\% of the total dust mass in this
object.  Such a cold reservoir may thus increase the total dust
content by a factor of a few, but likely not by an order of magnitude.
We thus conclude that the dust--to--gas ratio (M$_{\rm
dust}$/M$_{gas}$) in our sample dwarfs appears to be significantly
lower than what is found in spiral galaxies (which can not simply be
explained by a linear scaling of this ratio with metallicity, see also
the discussion in Draine et al.\ 2007).

\section{Summary}
\label{S4}

We present observations of warm dust and atomic gas in seven dwarf
irregular galaxies in the M\,81 group using data from both {\it SINGS}
and {\it THINGS}.  Five of the seven targets have been detected with
{\it Spitzer} out to 160\,\mm\ (the dwarfs with the lowest star
formation rates, M\,81\,dwA and DDO\,165, are non--detections).  As
molecular gas in these systems has yet to be detected (only likely
exception: IC~2574, {Leroy \etal\ 2005}\nocite{leroy05}) the {\it
  Spitzer} observations give a first glimpse of the nature of the
non--atomic ISM in these galaxies and provide important information to
design follow--up observations of the molecular gas phase (e.g., to
select regions of interest for pointed observations using millimeter
telescopes).

We find that the warm dust emission as traced via the 70\,\mm\
observations is associated with high \HI\ column densities (N$_{\rm
  HI}\sim1\times10^{21}$\,cm$^{-2}$), close to the `canonical' star
formation threshold found by previous studies (e.g., {Skillman
  1996}\nocite{skillman96conf}, {Walter \& Brinks
  1999}\nocite{walter99}, {Schaye 2004}\nocite{schaye04}). Most
regions with \HI\ column densities of N$_{\rm
  HI}>2.5\times10^{21}$\,cm$^{-2}$ have dust emission associated with
them. For the brightest regions at 70\,\mm\ there is a good
correlation with the location of HII regions, indicating that active
star formation is needed to heat up the dust locally. However, in some
cases, there is diffuse dust emission present at larger radii which
does not appear to coincide with compact HII regions. This diffuse
emission is likely due to the re-processing of non-ionizing photons in
the ISM or the escape of radiation from the star formation regions
(see also Cannon et al.\ 2006b).

IRS spectroscopy in the brightest regions in IC~2574 and Holmberg~II
(which have comparable star formation rates) reveal distinctly
different spectral shapes: Whereas PAH features are clearly detected
in the spectrum of IC~2574 those features are weaker in Holmberg~II
(which has lower metallicity) by an order of magnitude.  This
emphazises that the strength of PAH features is not a simple linear
function of metallicity (see also Smith et al.\ 2006). The
PAH--to--TIR continuum ratio in IC~2574 is a factor of $\sim$7 less
than what is found in a typical SINGS spiral (a factor of $>20$ in the
case of Holmberg~II).

While the \HI\ masses are well-constrained, it is difficult to
constrain the dust masses with a high degree of certainty.  We
estimate dust masses of $\sim$10$^4$--$10^{6}$ \msun\ for individual
targets, resulting in an average dust--to--gas ratio (M$_{\rm
dust}$/M$_{\rm HI}$) of $\sim3\times10^{-4}$ ($1.5\times 10^{-3}$ if
only the \HI\ that is associated with dust emission is considered).
This can be compared to the range in values derived for the SINGS
galaxies by Draine \etal\ (2007, in prep.) of 0.005$<$M$_{\rm
dust}$/M$_{\rm gas}<$0.02. Thus, the average dust--to--gas ratio in
the dwarfs is lower by about an order of magnitude as compared to more
massive spirals (a finding that can not simply be explained by a
linear scaling with metallicity, see also Draine \etal\ 2007). Future
sensitive observations at longer (sub--mm) wavelengths are critical to
constrain the possible presence of a colder dust component in the
galaxies (not heated by the stellar population) that may be missed by
the MIPS observations.

We also find that the dwarf galaxies in our sample are underluminous
at 70\,\mm\ for a given SFR by about a factor of $\sim2$ compared to the
more massive and metal--rich galaxies in SINGS.  However,
interestingly, the average 70\,\mm/160\,\mm\ ratio in the dwarfs is
higher (factor of $\sim$2) than in the spiral galaxies.  In a
simplistic picture, this can be attributed to higher effective dust
temperatures in the dwarf galaxies (which shifts the peak of the warm
dust SED toward 70\,mm, cf. {Dale \etal\ 2005}\nocite{dale05}).
Similar conclusions on the dust temperature have been derived by other
authors studying more luminous dwarf systems ({Hunter \etal\
  1989}\nocite{hunter89}, {Dale \etal\ 2005}\nocite{dale05},
{Engelbracht \etal\ 2005}\nocite{engelbracht05}, {Cannon \etal\ 2005,
  2006a, 2006b}\nocite{cannon05,cannon06a,cannon06b}) and have been
explained in the context of stronger radiation fields in the dwarfs.
It is interesting to note that, if the dwarf galaxies had the same
temperature as the more massive spirals, the 70\mm\ luminosity for a
given SFR would decrease further (relative to the spirals). Overall,
there is a better correlation between the SFR (or optical magnitudes)
and the 70\,\mm\ luminosity than between the \HI\ mass and L(70\,\mm).
This provides additional evidence that the FIR emission in the sample
dwarf galaxies is powered by ongoing star formation and does not
strongly depend on the total \HI\ mass of the galaxy host.

\acknowledgements

Some of the data presented here are part of the {\it {\it Spitzer}
  Space Telescope} Legacy Science Program ``The {\it Spitzer} Nearby
Galaxies Survey (SINGS)'', which was made possible by NASA through
contract 1224769 issued by JPL/Caltech under NASA contract 1407. This
work was also supported in part by NSF grant AST--0406883.



\begin{thebibliography}{}

\bibitem[Asplund et al.(2005)]{2005ASPC..336...25A} Asplund, M., Grevesse, 
N., \& Sauval, A.~J.\ 2005, ASP Conf.~Ser.~336: Cosmic Abundances as 
Records of Stellar Evolution and Nucleosynthesis, 336, 25 

\bibitem[Barone \etal(2000)]{barone00} Barone, L.~T., Heithausen, A.,
H{\"u}ttemeister, S., Fritz, T., \& Klein, U.\ 2000, \mnras, 317, 649

\bibitem[Bendo et al.(2006)]{1208} Bendo, G.~J., et al.\ 2006, \apj, 652,
  283

\bibitem[Calzetti \etal(2005)]{calzetti05} Calzetti, D., \etal\ 2005, \apj,
633, 871

Calzetti et al.\, 2006, in prep.

\bibitem[Cannon \etal(2005)]{cannon05} Cannon, J.~M., \etal\ 2005, \apjl,
630, L37
 
\bibitem[Cannon et al.(2006a)]{cannon06a} Cannon, J.~M., et al.\ 
2006a, \apj, 647, 293 

\bibitem[Cannon et al.(2006b)]{2006ApJ...652.1170C} Cannon, J.~M., et al.\ 
2006, \apj, 652, 1170 

\bibitem[Dale \etal(2001)]{dale01} Dale, D.~A., Helou, G., Contursi, A.,
Silbermann, N.~A., \& Kolhatkar, S.\ 2001, \apj, 549, 215

\bibitem[Dale \& Helou(2002)]{dale02} Dale, D.~A., \& Helou, G.\ 2002, \apj,
576, 159

\bibitem[Dale \etal(2005)]{dale05} Dale, D.~A., \etal\ 2005, \apj, 633, 857

Dale, D.~A., \etal\ 2006, in prep.

\bibitem[Draine \& Li(2001)]{draine01} Draine, B.~T., \& Li, A.\ 2001, \apj, 
551, 807

\bibitem[Draine \& Li(2006)]{draine06} Draine, B., \& Li, A., 2006,
  ApJ, subm. (astro-ph/0608003)

\bibitem[Draine et al.(2007)]{draine07} Draine \etal\ 2007, in prep.

\bibitem[de Vries \etal(1987)]{devries87} de Vries, H.~W., Thaddeus, P., \&
Heithausen, A.\ 1987, \apj, 319, 723

\bibitem[Engelbracht \etal(2004)]{engelbracht04} Engelbracht, C.~W., \etal\
2004, \apjs, 154, 248

\bibitem[Engelbracht \etal(2005)]{engelbracht05} Engelbracht, C.~W., Gordon,
K.~D., Rieke, G.~H., Werner, M.~W., Dale, D.~A., \& Latter, W.~B.\ 2005,
\apjl, 628, L29

\bibitem[Gallagher \etal(1991)]{gallagher91} Gallagher, J.~S., Hunter,
D.~A., Gillett, F.~C., \& Rice, W.~L.\ 1991, \apj, 371, 142

\bibitem[Gordon \etal(2005)]{gordon05} Gordon, K.D., \etal\ 2005, \pasp, 
177, 503

\bibitem[Haas \etal(1998)]{haas98} Haas, M., Lemke, D., Stickel, M.,
Hippelein, H., Kunkel, M., Herbstmeier, U., \& Mattila, K.\ 1998,
\aap, 338, L33

\bibitem[Hinz et al.(2006)]{2006astro.ph..7265H} Hinz, J.~L., Misselt, K., 
Rieke, M.~J., Rieke, G.~H., Smith, P.~S., Blaylock, M., \& Gordon, K.~D.\ 
2006, ArXiv Astrophysics e-prints, arXiv:astro-ph/0607265 


\bibitem[Hippelein \etal(2003)]{hippelein03} Hippelein, H., Haas, M.,
Tuffs, R.~J., Lemke, D., Stickel, M., Klaas, U., V{\" o}lk, H.~J.\
2003, \aap, 407, 137

\bibitem[Houck et al.(2004)]{2004ApJS..154...18H} Houck, J.~R., et al.\ 
2004, \apjs, 154, 18 

\bibitem[Houck \etal(2004)]{houck04} Houck, J.~R., \etal\ 2004, \apjs, 154, 211

\bibitem[Hunter \etal(1989)]{hunter89} Hunter, D.~A., Gallagher, J.~S., Rice,
W.~L., \& Gillett, F.~C.\ 1989, \apj, 336, 152

\bibitem[Hunter \etal(2001)]{hunter01} Hunter, D.~A., \etal\ 2001, \apj, 553,
121

\bibitem[Jackson et al.(2006)]{2006ApJ...646..192J} Jackson, D.~C., Cannon, 
J.~M., Skillman, E.~D., Lee, H., Gehrz, R.~D., Woodward, C.~E., \& 
Polomski, E.\ 2006, \apj, 646, 192 

\bibitem[Jorsater \& van Moorsel(1995)]{jorsater95} J\"ors\"ater, S., \& van
Moorsel, G.~A.\ 1995, \aj, 110, 2037

\bibitem[Karachentsev \etal(2002)]{karachentsev02} Karachentsev, I.~D., et
al.\ 2002, \aap, 383, 125

\bibitem[Karachentsev et al.(2003)]{2003A&A...398..479K} Karachentsev, 
I.~D., et al.\ 2003, \aap, 398, 479 

\bibitem[Kennicutt(1998)]{kennicutt98} Kennicutt, R.~C.\ 1998, \araa, 36, 189

\bibitem[Kennicutt \etal(2003)]{kennicutt03} Kennicutt, R.~C., \etal\ 2003, 
\pasp, 115, 928

Kennicutt, R.~C. et al. 2006, ApJS, in prep.

\bibitem[Kobulnicky \& Skillman(1997)]{1997ApJ...489..636K} Kobulnicky, 
H.~A., \& Skillman, E.~D.\ 1997, \apj, 489, 636 

\bibitem[Kobulnicky \& Skillman(1996)]{1996ApJ...471..211K} Kobulnicky, 
H.~A., \& Skillman, E.~D.\ 1996, \apj, 471, 211 

Lee, J.C., 2006, PhD Thesis, University of Arizona

\bibitem[Leroy \etal(2005)]{leroy05} Leroy, A., Bolatto, A.~D., Simon, J.~D., 
\& Blitz, L.\ 2005, \apj, 625, 763 

\bibitem[Li \& Draine(2001)]{li01} Li, A., \& Draine, B.~T.\ 2001, \apj, 
554, 778

\bibitem[Li \& Draine(2002)]{li02} Li, A., \& Draine, B.~T.\ 2002, \apj, 
576, 762

\bibitem[Lu et al.(2003)]{2003ApJ...588..199L} Lu, N., et al.\ 2003, \apj, 
588, 199 

\bibitem[Galliano et al.(2003)]{2003A&A...407..159G} Galliano, F., Madden, 
S.~C., Jones, A.~P., Wilson, C.~D., Bernard, J.-P., \& Le Peintre, F.\ 
2003, \aap, 407, 159 

\bibitem[Madden et al.(2006)]{2006A&A...446..877M} Madden, S.~C.,
  Galliano, F., Jones, A.~P., \& Sauvage, M.\ 2006, \aap, 446, 877

\bibitem[Melisse \& Israel(1994)]{melisse94a} Melisse, J.~P.~M., 
\& Israel, F.~P.\ 1994a, \aap, 285, 51 

\bibitem[Melisse \& Israel(1994)]{melisse94b} Melisse, J.~P.~M., 
\& Israel, F.~P.\ 1994b, \aaps, 103, 391 

\bibitem[Miller \& Hodge(1994)]{miller94} Miller, B.~W., \& 
Hodge, P.\ 1994, \apj, 427, 656

Moustakas, J., et al.\ 2007, in prep.

\bibitem[Pilyugin \& Thuan(2005)]{2005ApJ...631..231P} Pilyugin,
  L.~S., \& Thuan, T.~X.\ 2005, \apj, 631, 231

\bibitem[O'Halloran et al.(2006)]{2006ApJ...641..795O} O'Halloran, B., 
Satyapal, S., \& Dudik, R.~P.\ 2006, \apj, 641, 795 

\bibitem[Ott \etal(2001)]{ott01} Ott, J., Walter, F., Brinks, E., Van Dyk,
S.~D., Dirsch, B., \& Klein, U.\ 2001, \aj, 122, 3070

\bibitem[Popescu \etal(2002)]{popescu02} Popescu, C.~C., Tuffs, R.~J.,
V{\"o}lk, H.~J., Pierini, D., \& Madore, B.~F.\ 2002, \apj, 567, 221

\bibitem[Popescu \& Tuffs(2003)]{2003A&A...410L..21P} Popescu, C.~C., \& 
Tuffs, R.~J.\ 2003, \aap, 410, L21 

\bibitem[Puche \etal(1992)]{puche92} Puche, D., Westpfahl, D., Brinks, E., \&
Roy, J.-R.\ 1992, \aj, 103, 1841

\bibitem[Rigby \& Rieke(2004)]{2004ApJ...606..237R} Rigby, J.~R., \& Rieke, 
G.~H.\ 2004, \apj, 606, 237

\bibitem[Rosenberg et al.(2006)]{2006ApJ...636..742R} Rosenberg, J.~L., 
Ashby, M.~L.~N., Salzer, J.~J., \& Huang, J.-S.\ 2006, \apj, 636, 742 

\bibitem[Salpeter(1955)]{1955ApJ...121..161S} Salpeter, E.~E.\ 1955, \apj, 
121, 161 


\bibitem[Schaye(2004)]{schaye04} Schaye, J.\ 2004, \apj, 609, 
667

\bibitem[Skillman \& Bothun(1986)]{skillman86} Skillman, E.~D., \&
Bothun, G.~D.\ 1986, \aap, 165, 45

\bibitem[Skillman(1996)]{skillman96conf} Skillman, E.~D.\ 1996, ASP 
Conf.~Ser.~106: The Minnesota Lectures on Extragalactic Neutral Hydrogen, 
106, 208 

\bibitem[Smith et al.(2006)]{2006astro.ph.10913S} Smith, J.~D.~T., et al.\ 
2006, ArXiv Astrophysics e-prints, arXiv:astro-ph/0610913 




\bibitem[Taylor \etal(1998)]{taylor98} Taylor, C.~L., Kobulnicky, H.~A., \&
Skillman, E.~D.\ 1998, \aj, 116, 2746

\bibitem[Telesco(1988)]{1988ARA&A..26..343T} Telesco, C.~M.\ 1988, \araa, 
26, 343 

\bibitem[Thronson \& Telesco(1986)]{thronson86} Thronson, H.~A., 
Jr., \& Telesco, C.~M.\ 1986, \apj, 311, 98 


\bibitem[Walter \& Brinks(1999)]{walter99} Walter, F., \& Brinks, E.\ 1999,
\aj, 118, 273

\bibitem[Walter \etal(2005)]{walter05conf} Walter, F., Brinks, E., de Blok,
W.~J.~G., Thornley, M.~D., \& Kennicutt, R.~C.\ 2005, ASP Conf.~Ser.~331:
Extra-Planar Gas, 331, 269


\bibitem[Wu et al.(2006)]{2006ApJ...639..157W} Wu, Y., Charmandaris,
  V., Hao, L., Brandl, B.~R., Bernard-Salas, J., Spoon, H.~W.~W., \&
  Houck, J.~R.\ 2006, \apj, 639, 157


\end{thebibliography}
\end{document}